\documentclass[aps,prb,twocolumn,superscriptaddress,longbibliography]{revtex4-2}
\usepackage{graphicx}
\usepackage{amsmath}
\usepackage{amssymb}
\usepackage{mathtools}
\usepackage{xcolor}
\usepackage{braket}
\usepackage[colorlinks=true, citecolor=blue, urlcolor=blue, linkcolor=blue,bookmarks=false,hypertexnames=true]{hyperref} 
\usepackage{appendix}
\usepackage{array}
\usepackage[column=O]{cellspace}
\newcolumntype{P}[1]{>{\centering\arraybackslash}p{#1}}
\begin{document}
\graphicspath{}
\preprint{APS/123-QED}
\title{Atomistic Approach to Exciton-Phonon Couplings in Semiconductor Quantum Dots}
\date{\today}
\author{Yasser Saleem}
\affiliation{Condensed Matter Theory, Department of Physics, TU Dortmund, 44221 Dortmund, Germany.}
\affiliation{
Institute for Theoretical Physics and Astrophysics,
and Würzburg-Dresden Cluster of Excellence on Complexity, Topology and Dynamics in Quantum Matter ctd.qmat,
Julius-Maximilians-Universität Würzburg, Am Hubland, D-97074 Würzburg, Germany}
\author{Moritz Cygorek}
\affiliation{Condensed Matter Theory, Department of Physics, TU Dortmund, 44221 Dortmund, Germany.}
\begin{abstract}
\noindent We present a fully atomistic approach to exciton-phonon coupling in semiconductor quantum dots that bridges microscopic electronic-structure calculations with non-Markovian open-quantum-system dynamics. On the example of an InAsP quantum dot embedded in an InP matrix, we compute single-particle states using an \emph{ab initio}-parametrized tight-binding model, then obtain correlated many-body wave functions of neutral excitons, biexcitons, and charged trions via the configuration-interaction method. Using these correlated states, we compute the exciton-phonon coupling matrix elements. The resulting  phonon spectral densities for different excitonic complexes are compared with the widely used analytical super-Ohmic form and reveal  deviations at higher energies originating from the realistic dot geometry and atomistic wave functions, whereas configuration mixing is found to play only a minor role. Furthermore, we extract radiative lifetimes comparable to values experimentally measured. As a direct application, we simulate the emission brightness of a pulsed-driven quantum dot and demonstrate that the atomistically derived spectral density substantially broadens the region of efficient off-resonant excitation compared to the analytical model. The presented framework provides a nearly parameter-free route to simulate the non-Markovian open quantum dynamics in semiconductor quantum dots.

\vspace{6mm}
Corresponding author email: yassersaleem461@gmail.com
\end{abstract}
\maketitle
\section{Introduction} 
Semiconductor quantum dots (QDs) are promising building blocks for quantum technologies, particularly because of their strong interaction with light and their compatibility with photonic structures such as waveguides and microcavities\cite{ota2011spontaneous,schwagmann2011chipSinglePhotoWavefuide,del2011generationMicroCavity,schumacher2012cavity,qian2018two,huber2020filter,liu2024dynamic,liu2025quantumNAture}. A plethora of applications exists, including quantum-dot lasers\cite{Arakawa1982SemiconductingLasers,fafard1996redlaser,bimberg2002ingaasLaser}, single-photon emitters\cite{schwagmann2011chipSinglePhotoWavefuide,senellart2017highSinglePhoton,huber2017highly,schweickert2018demandsinglePhoton,PRLCosacchiBrightness,huber2020filter,arakawa2020progressSinglePhoton}, sources of entangled photon pairs\cite{schumacher2012cavity,huber2018semiconductorEntangledPhotonPairs}, and generators of highly entangled photon cluster states\cite{Lindner2009PhotonCluster,schwartz2016deterministicClusterStates}. 

After steady technological advances over a few decades, QD-based nonclassical light sources now show remarkable figures of merit, such as single-photon purities of  $g^{(2)}<0.002$ \cite{huber2017highly,schweickert2018demandsinglePhoton} as well as entanglement fidelities of 99.4\%\cite{liu2025quantumNAture}. This naturally raises the question what currently limits their performance, and what can be done to further bring them close to perfection. Some issues can be addressed by controlling the external laser driving. For example, imperfect state-preparation fidelity can be addressed by robust protocol relying on adiabatic rapid passage~\cite{ARPWu2011,ARPSimon2011}, phonon-assisted state preparation~\cite{PRLCosacchiBrightness,Groll_2021}, or single-shot Floquet driving~\cite{floquet}. Spectral separation between emitted photons and the exciting laser can be ensured by coherent two-photon excitation~\cite{SUPER}, and also the unintended Stark shift due the excitation pulse that negatively affects the entanglement fidelity of emitted photon pairs~\cite{SeidelmannLimit} can be remedied by appropriate driving protocols~\cite{Heinisch2024,BrachtOpticaQuantum2023}. Key for the development of advanced driving protocol has been the emergence of numerically exact techniques for simulating the dynamics in QDs and QD-cavity systems~\cite{ACE,ACE_code,DnC}, which fully account for the non-Markovian feedback of the strongly coupled longitudinal acoustic (LA) phonon environment.

Other issues have a structural origin. This includes the finestructure splittings of excitonic states, which originates from the asymmetry of excitonic wave functions~\cite{FSS_Takagahara} and spoils entanglement fidelities~\cite{SeidelmannStrongToWeak}. Moreover, thermal activation of higher excited states in relatively large QDs limits the photon indistinguishability already at moderate temperatures~\cite{MuljarovZimmermann}. Two-photon emission~\cite{SeidelmannStrongToWeak} as well as Floquet-driving~\cite{floquet} crucially depend on the value of the biexciton binding energy, while the shape of the phonon spectral density is important for understanding the phonon sidebands in spectra as well as for the performance of adiabatic excitation protocols~\cite{Kaldewey2017}.

\begin{figure*}[ht]    \includegraphics[width=\linewidth]{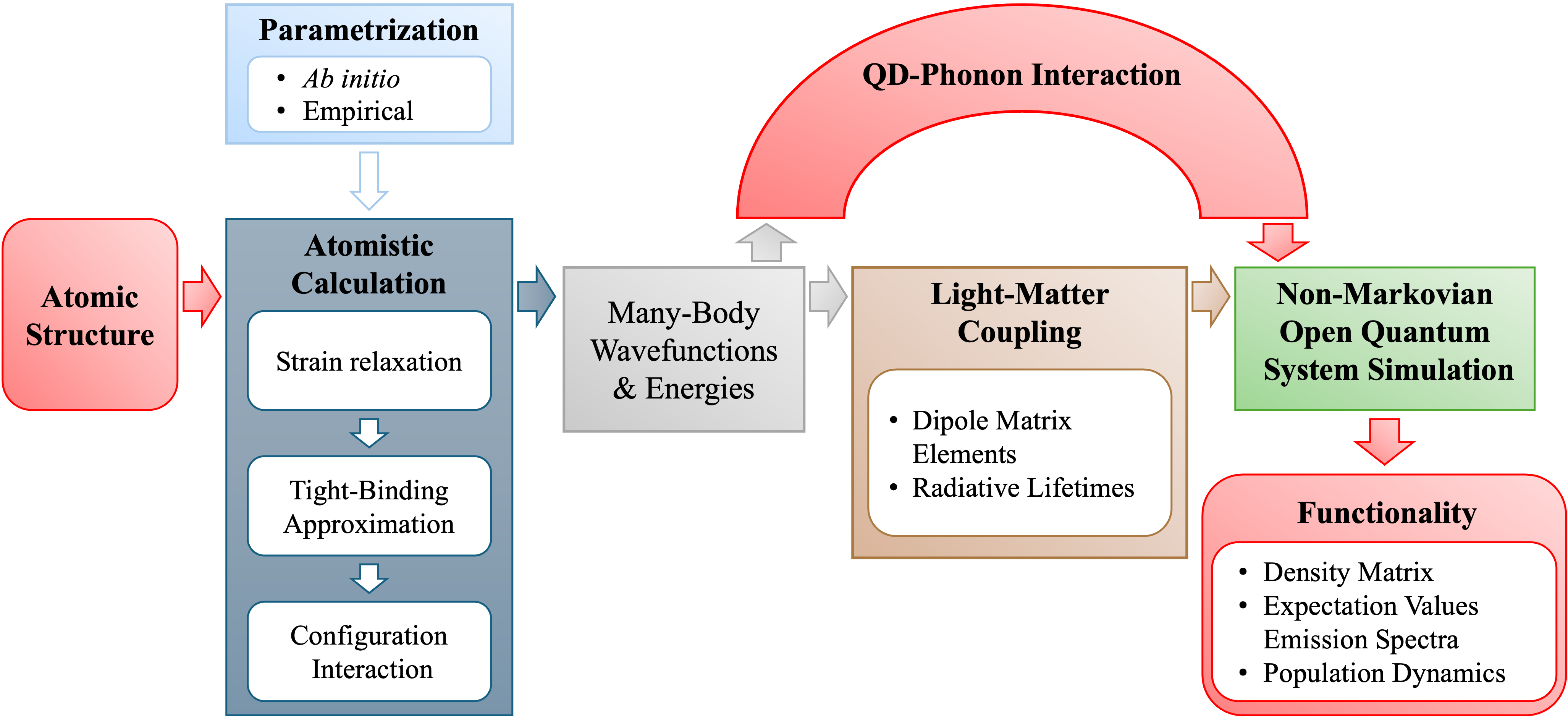}
    \caption{ Schematic workflow of the theoretical framework. Starting from the atomic structure, parametrized either by \emph{ab initio} or empirical methods, strain relaxation is followed by atomistic tight-binding calculations and configuration interaction to obtain many-body wavefunctions and energies. These serve as input for the evaluation of light–matter coupling (dipole matrix elements, radiative lifetimes) and for modeling the quantum-dot–phonon interaction. The resulting information enters a non-Markovian open quantum systems simulation, which can provide the time-dependent density matrix, expectation values, emission spectra, and population dynamics.  
} 
\label{fig:fig1}
 \end{figure*}

The intrinsic randomness of shape and size of self-assembled QDs made it necessary to search, select, and tune individual QDs with the desired electronic and optical properties. Deterministic growth techniques~\cite{dalacu2009selective,dalacu2012ultraclean,dalacu2021tailoring,chen2016controlling,sallen2009exciton}, such as VLS growth of QDs in nanowires now enable more direct control of QD geometries. How the geometry should be chosen to result in desired properties can be simulated atomistically using density-functional-theory-informed million-atom tight-binding simulations in combination with configuration interaction methods for many-body complexes~\cite{zielinski2010atomistic,cygorek202Atomistic,CIPSI, Laferriere2021} [see Fig.~\ref{fig:fig1}]. In general, atomistic methods have been used to predict single-particle levels, fine-structure splittings, excitonic binding energies and lifetimes, tunability  and excitonic wavefunctions with high accuracy\cite{hawrylak1996electronic,zunger1998electronic,bester2003pseudopotential,jaskolski2006strain,klimeck2007atomisticI,klimeck2007atomisticII,NarvaezBesterZungerPRB2005,zielinski2010atomistic,cygorek202Atomistic,zielinski2025double,zielinski2013fine,swiderski2017atomistic,swiderski2019electric,swiderski2021electric}. So far not much attention has been given to the derivation of QD-phonon couplings from atomistic principles.

Combining both approaches---electronic structure calculation and non-Markovian open quantum systems dynamics---should thus enable realistic and accurate predictions of how changes in the atomistic details of semiconductor quantum nanodevices affect their performance, and what combination of growth and excitation parameters can yield near-perfect quantum devices. The full pipeline for simulations from structure to functionality is sketched in Fig.~\ref{fig:fig1}. While most elements of this pipeline have already been developed, a central link, which has so far rarely been explored~\cite{Krzykowski2020} and thus requires a special focus, is the determination of the interactions between excitonic complexes in QDs from microscopic principle.

In this article, we complete this pipeline. We focus on an InAsP QD embedded in an InP matrix. Starting from an \emph{ab initio}–based tight-binding model, we compute the single-particle states and then apply the configuration-interaction method to obtain correlated optical complexes, including excitons, trions, and biexcitons. From these many-body states we determine the dipole matrix elements and extract the corresponding radiative lifetimes. Most importantly, we compute the coupling of these correlated complexes to phonons and thereby obtain their phonon spectral densities. These atomistically derived spectral densities—which incorporate configuration mixing and the realistic dot geometry—are compared with the widely used analytical model that lacks these microscopic features. To illustrate the consequences of these differences, we simulate driving the InAsP/InP QD with a Gaussian pulse and compute the resulting emission brightness as a function of pulse area and detuning.

\section{Theoretical Model}
In this section, we outline the theoretical framework that connects atomistic simulations of the quantum-dot structure to the full non-Markovian open quantum system dynamics. The overall workflow is summarized in Fig.~\ref{fig:fig1}, and the following subsections discuss each stage in detail. Starting from the atomistic structure of the dot, we perform strain relaxation and atomistic tight-binding calculations using parameters obtained either from \emph{ab initio} fits or empirical models. The resulting single-particle states form the basis for configuration-interaction calculations, from which we obtain many-body wavefunctions, light–matter coupling elements, and exciton–phonon matrix elements. These microscopic ingredients then feed into a non-Markovian open-quantum-system simulation that captures the time-dependent dynamical response of the QD. 

\subsection{Open Quantum Systems}
Before examining the individual stages of the workflow in Fig.~\ref{fig:fig1}, we begin by outlining the final problem we seek to address: the non-Markovian open quantum system dynamics of the QD. This formulation determines the observables of interest and highlights which microscopic ingredients---many-body states, dipole matrix elements, and exciton–phonon couplings---must be obtained from the preceding atomistic and many-body calculations.

First, let us consider a driven semiconductor QD as an open quantum system. 
The Hamiltonian is generally written as a sum of a system Hamiltonian $H_S(t)$, 
an environment $H_E(t)$, and loss terms that enter via Lindblad operators. 
In this work the system Hamiltonian describes the interaction of the QD with light. 
For example, if we restrict to the ground state and the lowest-energy exciton state $X$, 
the QD forms a driven two-level system with Hamiltonian
\begin{equation}
    H_S(t) = -\hbar\delta \ket{X}\bra{X} 
    + \hbar\Omega(t)\bigl(\sigma_- + \sigma_+ \bigr),
    \label{eq:Driven2lsysH}
\end{equation}
where $\sigma_+=\ket{X}\bra{GS}$ and $\sigma_-=\ket{GS}\bra{X}$ are raising and lowering operators. Here, $\delta$ is the detuning of the light field, and $\hbar\Omega(t)=-e\braket{X|\textbf{E}_0(t)\cdot \textbf{r}|GS}$ is the Rabi driving. Dissipation is included through Lindblad operators that account for spontaneous emission. On the example of the two level system this can be written as
\begin{equation}
    \mathcal{L}_{X}\left[\rho_s\right] = \gamma_{X}\left[\sigma_-\rho_s\sigma_+-\frac{1}{2}\left(\sigma_+\sigma_-\rho_s+\rho_s\sigma_+\sigma_-\right)\right].
    \label{eq: LindbladOp}
\end{equation}
where $\gamma_{X}$ is the spontaneous emission rate from $X$ to the ground state, and $\rho_s$ is the reduced density matrix of the system. The environment, consisting of phonons, is described by the generalized spin-boson Hamiltonian
\begin{equation}
H_E = \sum_\textbf{k} \hbar \omega_\textbf{k} b_\textbf{k}^\dagger b_\textbf{k} + \sum_\textbf{k} \hbar \left( 
\Bar{g}_\textbf{k}^{X*} b_\textbf{k}^\dagger + \Bar{g}^X_\textbf{k} b_\textbf{k} \right) \hat{O},
\label{eq: GeneralizedBose}
\end{equation}
with bosonic creation and annihilation operators $b_\mathbf{k}^\dagger$ and $b_\mathbf{k}$, mode frequencies $\omega_\mathbf{k}$. 
For a two-level system coupled to a bath, e.g., of longitudinal acoustic phonons, the interaction leads to dephasing, which is generally described~\cite{mahanbook} by a system operator $\hat{O}=\ket{X}\bra{X}$, where $\Bar{g}_\mathbf{k}^X$ denotes the coupling constant between phonon mode $\mathbf{k}$ and the exciton state $\ket{X}$ measured relative to the $GS$. Explicitly,  $\Bar{g}_\mathbf{k}^X=g_\mathbf{k}^X -g_\mathbf{k}^{GS}$, where are the $g_\mathbf{k}^{X/GS}$ are the coupling constants between a phonon mode $\textbf{k}$ and the $X/GS$. 
The environment is conventionally characterized by the phonon spectral density,
\begin{equation}
    J(\omega) = \sum_\mathbf{k} \left|g^X_\mathbf{k}-g^{GS}_\mathbf{k}\right|^2 
    \delta\!\left(\omega-\omega_\mathbf{k}\right).
\end{equation}
If, as is common practice~\cite{Krummheuer2002PRB,Nazir_2016,Vagov2002PRB, WiggerLA2020}, one assumes excitonic wavefunctions to be made up of products of Gaussian functions for electrons and holes separately and the coupling to be dominated by acoustic (LA) phonons, the following analytical spectral density can be derived:
\begin{equation}
J_{a}(\omega) = \frac{\omega^3}{4 \pi^2 \rho \hbar c_s^5} 
\left( 
D_e e^{-\omega^2 a_e^2 / (4 c_s^2)} - D_h e^{-\omega^2 a_h^2 / (4 c_s^2)} 
\right)^2,
\label{eq: AnalyticalSpectralDensity}
\end{equation}
where $\rho$ is the mass density, $c_s$ is the speed of LA phonons in the material, $a_e$, $a_h$ are electron and hole confinement lengths, and $D_h$, $D_e$ are electron and hole deformation potential constants. Throughout this paper, we take $c_s=5110$ m/s, $\rho=4.7$ $\mathrm{g/cm^{3}}$ , $D_c=7$ eV, and $D_v$ = -3.5 eV~\cite{Krummheuer2002PRB,Krummheuer2005PRB,Nazir_2016}. Although Eq.~\eqref{eq: AnalyticalSpectralDensity} is derived from microscopic considerations, a set of assumptions was made in deriving this form of the phonon spectral density. This form of the phonon spectral density does not account for the atomistic resolution of the electronic wavefunctions, the effects of correlations on the exciton state, and assumes a spherical geometry of the QD. Nevertheless it is argued in Ref.~\cite{Reiter2017} that using $a_e$ and $a_h$ as free fitting parameters gives $J_a(\omega)$ the flexibility to also approximate the spectral density for nonspherical QDs. We will refer to this analytical form of the phonon spectral density throughout the paper as a form of comparison with our atomistic results.

The states entering this two-level model originate from the underlying interacting many-body eigenstates of the system. Even in the two-level approximation, the Hamiltonians above contain a wide range of parameters that must be obtained either experimentally or through microscopic calculations. 
In this work we follow an atomistic approach: we compute the dipole matrix elements determining the light–matter coupling, extract Lindblad rates by evaluating the lifetimes of excitonic complexes, and, most importantly, determine the coupling of many-body complexes (excitons, trions, and biexcitons) to the bulk phonon environment.
\subsection{Tight-binding model}
We begin by briefly discussing the TB model which serves to obtain the single-particle states of the QD. A more detailed analysis can be found in previous work~\cite{cygorek202Atomistic}. The TB Hamiltonian can be written as
\begin{equation}
\begin{split}
    H_{TB} = &\sum_{i} \sum_{\alpha} \epsilon_{\alpha i} c_{\alpha i}^\dagger c_{\alpha i}
+ \sum_{i} \sum_{\alpha,\beta} \lambda_{\alpha\beta,i} c_{\alpha i}^\dagger c_{\beta i} \\
&+ \sum_{i,j}  \sum_{\alpha,\beta} t_{\alpha i,\beta j} c_{\alpha i}^\dagger c_{\beta j},
\end{split}
\label{eq: TBHam}
\end{equation}
where $c^{\dagger}_{\alpha i}/c_{\alpha i}$ create or annihilate an electron on orbital $\alpha$ localized on site $i$, $\epsilon_{\alpha i}$ is the onsite energy, $\lambda_{\alpha\beta,i}$ describes the spin-orbit interaction and  $t_{\alpha i,\beta j}$
are the hopping terms. Here, we employ an spds$^*$ model with 20 orbitals per atom. The tight-binding parameters, which include strain corrections, are obtained by fitting to the DFT band structures with the band gap shifted to reproduce the empirical one. Diagonalizing the million-atom tight-binding Hamiltonian in Eq.~\eqref{eq: TBHam} and extracting states close to the band gap, the $n$-th single-particle wavefunction localized in the QD region is given as a linear combination of atomic orbitals:
\begin{equation}
    \psi_n(\textbf{r}) = \sum_{\alpha}\sum_{i}C^n_{\alpha i}\phi_{\alpha}\left(\textbf{r}-\textbf{R}^0_i\right),  
    \label{eq: TBWF}
\end{equation}
where $\textbf{R}^0_i$ is the equilibrium position of atom $i$, $\phi_{\alpha}\left(\textbf{r}-\textbf{R}^0_i\right)$ is an atomic orbital localized on site $i$, and $C^n_{\alpha i}$ are expansion coefficients. 
\subsection{Configuration interaction}
The elementary electronic excitation of QDs are excitonic complex, such as exciton ($X$), trions ($X^-,X^+$), or biexcitons ($XX$). To obtain the many-body wave functions, energies, spin states, and dipoles, one has to solve the many-body problem including Coulomb interaction, e.g., using the configuration interaction method.

We start with the many-body Hamiltonian in the electron-hole language as~\cite{cygorek202Atomistic,FranzCarl2025}
\begin{equation}
\begin{split}
H_{\text{MB}} &= \sum_i E_i^{(e)} d_i^\dagger d_i + \sum_p E_p^{(h)} h_p^\dagger h_p \\
&\quad + \frac{1}{2} \sum_{ijkl} \langle ij|V_{ee}|kl\rangle d_i^\dagger d_j^\dagger d_k d_l \\
&\quad + \frac{1}{2} \sum_{pqrs} \langle pq|V_{hh}|rs\rangle h_p^\dagger h_q^\dagger h_r h_s \\
&\quad - \sum_{i q r l} \left( \langle iq|V^{\text{dir}}_{eh}|rl\rangle 
- \langle iq|V^{\text{exc}}_{eh}|lr\rangle \right) d_i^\dagger h_q^\dagger h_r d_l.
\end{split}
\label{eq:MBHam}
\end{equation}
where now $d_i^\dagger/d_i$ creates/annihilates an electron on single-particle state $i$ and $h_p^\dagger/h_p$ creates/annihilates a hole on single-particle state $p$. 
The single-particle energies for electrons $E^{(e)}_i$ and holes $E^{(h)}_p$ are determined by the eigenvalues of the tight-binding Hamiltonian Eq.~\eqref{eq: TBHam}. Coulomb matrix elements are calculated from the respective wave functions, where we consider the direct electron-electron interaction
$\langle ij|V_{ee}|kl\rangle$, the hole-hole interactions are defined by $\langle pq|V_{hh}|rs\rangle$, and the electron-hole direct and exchange interaction is given by $\langle iq|V^{\text{dir}}_{eh}|rl\rangle $, $\langle iq|V^{\text{exc}}_{eh}|lr\rangle$. 

The many-body wavefunction for a many-body state $\lambda$ can be expanded as
\begin{equation}
    \ket{\Psi_\lambda}=\sum_{e,h}A^\lambda_{\{e\},\{h\}}\prod_{q}^{N_h}h^{\dagger}_{h_q}\prod_{p}^{N_e}d^{\dagger}_{e_p}\ket{GS},
    \label{MB:WF}
\end{equation}
where $\ket{GS}$ corresponds to the charge neutral ground state and $e/h$ corresponds to a sum over the electron/hole basis states. The coefficients $A^\lambda_{\{e\},\{h\}}$ are obtained by diagonalizing the many-body Hamiltonian Eq.~\eqref{eq:MBHam} in the basis of configurations that form the many-body complex of interest. For example, in the case of the exciton there is a single-electron ($N_e=1$) and a single hole $N_h=1$, and thus we diagonalize Eq.~\eqref{eq:MBHam} expanded in configurations of the type $h^{\dagger}_{h_q}d^{\dagger}_{e_p}\ket{GS}$.

\subsection{Coupling to Phonons}
The methods outlined above follow well-established formulations. 
What has been missing in a full atomistic description of confined nanostructure---and the central focus of the present article---is the evaluation of QD-phonon couplings as needed in the environment Hamiltonian in Eq.~\eqref{eq: GeneralizedBose}.
Since we consider not only excitons but also general excitonic complexes, the coupling constant is obtained from the matrix element $\langle \Psi_\lambda|V_{e-p} | \Psi_\lambda \rangle$, where $ \lambda$ indexes the many-body excitonic complexes and
$V_{e-p}$ defines the electron-phonon interaction (in the harmonic approximation). $V_{e-p}$ is explicitly given by\cite{mahanbook,Nazir_2016}
\begin{equation}
    V_{e-p}  = -\sum_{m}\mathbf{Q}_{m}\cdot \nabla \left(v_{e-ion}\left(\mathbf{r}-\mathbf{R}^0_{m}\right)\right).
\end{equation}
where $\mathbf{Q}_{m}$ are small displacements of atom $m$ about its equilibrium position $\mathbf{R}^0_{m}$, and $v_{e-ion}\left(\mathbf{r}-\mathbf{R}^0_{m}\right)$ are electron-ion interactions. Now expanding $v_{e-ion}\left(\mathbf{r}-\mathbf{R}^0_{m}\right)$ in a Fourier series, and neglecting Umklapp processes we obtain 
\begin{equation}
 V_{e-p}  = -\frac{i}{N_{UC}}\sum_{\mathbf{k}\in BZ}\mathbf{Q}_{\mathbf{k}}\cdot \mathbf{k} e^{i\mathbf{k}\cdot\mathbf{r}} v(\mathbf{k}).
\end{equation}
where $N_{UC}$ is the number of unit cells, $v(\mathbf{k})$ are Fourier coefficients, $\mathbf{k}$ are phonon wavevectors contained in the first Brillouin zone (BZ), and $\mathbf{Q}_{\mathbf{k}}=\sum_m \mathbf{Q}_me^{i \mathbf{k}\cdot\mathbf{R}^0_m}$ are the normal modes. Modeling phonon normal modes as harmonic oscillators, they are quantized as usual by $Q_{\mathbf{k}}=i\sqrt\frac{\hbar}{{2M\omega_{\mathbf{k}}}}\left(b_{\mathbf{k}}+b^{\dagger}_{-\mathbf{k}}\right)\mathbf{e}_{\mathbf{k}}$, where $b_\mathbf{k}$ and $b^\dagger_\mathbf{k}$ are bosonic annihilation and creation operators, respectively, $M$ is the average mass of a unit cell, $\mathbf{e}_{\mathbf{k}}$ is the polarization direction, and $\omega_\mathbf{k}$ is the normal mode energy. We note neglecting Umklapp processes is a good approximation when considering only low-energy excitations. In this case, only low-energy phonons contribute, corresponding to small momentum transfer $\textbf{k}$, whereas umklapp processes require large momentum transfers.
In semiconductor QDs, the dominant contribution to low-energy excitations arises from longitudinal acoustic (LA) phonons \cite{Takagahara1993PRL,Vagov2002PRB,Krummheuer2002PRB,Nazir_2016,niehues2018strain,wigger2021resonance}.  
Thus, we confine ourselves to LA phonons, i.e., phonons with polarization parallel to $\mathbf{k}$ and with linear dispersion $\omega_{\mathbf{k}}=c_s|\mathbf{k}|$. 
Putting everything together, we obtain
\begin{equation}
 V_{e-p}  = \sum_{\mathbf{k}} M_{\mathbf{k}} e^{i\mathbf{k}\cdot\mathbf{r}} v(\mathbf{k})\left(b_{\mathbf{k}}+b^{\dagger}_{-\mathbf{k}}\right),
 \label{eq:Vep}
\end{equation}
where $M_{\mathbf{k}} = \sqrt{\tfrac{|\mathbf{k}|}{2M\hbar c_sN_{UC}}}$. 
The coupling matrix element is then
\begin{equation}
\braket{\Psi_\lambda|V_{e-p}|\Psi_\lambda}=\hbar\sum_{\mathbf{k}} g^{\lambda}_{\mathbf{k}}\left(b_{\mathbf{k}}+b^{\dagger}_{-\mathbf{k}}\right),
\label{eq:ephononcouplingsimple}
\end{equation}
with
\begin{equation}
g^{\lambda}_{\mathbf{k}}=M_{\mathbf{k}}\!\left[D_c\braket{\Psi_\lambda|\hat{\varrho}_e(\mathbf{k})|\Psi_\lambda}-D_v\braket{\Psi_\lambda|\hat{\varrho}_h(\mathbf{k})|\Psi_\lambda}\right],
\label{eq:CouplingConstantsg}
\end{equation}
where $g^{\lambda}_{\mathbf{k}}$ represents the coupling of a many-body excitonic state $\lambda$ to phonons. 
Here, we have approximated $v(\mathbf{k})$ to lowest order as a constant, which is experimentally determined and takes different values for conduction-band ($D_c$) and valence-band ($D_v$) states \cite{Nazir_2016,Krummheuer2002PRB, Vagov2002PRB}. 
The electron and hole form factors are defined as
\begin{equation}
    \hat{\varrho}_e(\mathbf{k}) = \sum_{n\in \text{CB}} \varrho_{n}(\mathbf{k}) d^{\dagger}_{n} d_{n},
    \label{eq:FFe}
\end{equation}
\begin{equation}
    \hat{\varrho}_h(\mathbf{k}) = \sum_{n\in \text{VB}} \varrho_{n}(\mathbf{k}) h^{\dagger}_{n} h_{n},
    \label{eq:FFh}
\end{equation}
and their calculation is described in Appendix~\ref{app:FF}. 
The phonon spectral density describing dephasing between states $\lambda$ and $\lambda'$ is then
\begin{equation}
J^{\lambda-\lambda'}(\omega)=\sum_{\mathbf{k}}\left|g^{\lambda}_{\mathbf{k}}-g^{\lambda'}_{\mathbf{k}}\right|^2\delta\!\left(\omega-\omega_{\mathbf{k}}\right).
\label{eq:PhononSpectralDensity}
\end{equation}

Note that, here, we only consider electron-phonon coupling diagonal in the many-body eigenbasis, whereas $\langle \Psi_{\lambda'}| V_{e-p} | \Psi_\lambda\rangle$ is generally non-zero also for $\lambda'\neq\lambda$. However, direct transitions between excitonic states due to these offdiagonal terms tend to be suppressed because the energy differences between these states of tens of meV are typically much larger than the relevant phonon energies. While the offdiagonal couplings may affect, e.g., relaxation of high-orbital hole states with small level spacing~\cite{orbitalhole}, a thorough analysis of the impact of offdiagonal coupling terms is beyond the scope of this article.

\subsection{Spontaneous emission rates}
The spontaneous emission rates for semiconductor QDs can be extracted from Fermi's golden rule and are given by~\cite{Scully1997,NarvaezBesterZungerPRB2005}
\begin{equation}
\gamma_{i\rightarrow f} = \frac{n(\omega) \, \omega_{i\rightarrow f}^3 \, |\mathbf{d}_{i\rightarrow f}|^2}{3 \pi \epsilon_0 \hbar c^3},
    \label{eq:SERate}
\end{equation}
where $n(\omega)$ is the host material refractive index, $\omega_{i\rightarrow f}$ is the transition frequency, $c$ is the speed of light, and $\epsilon_0$ is the permittivity of free space. 
The dipole matrix element between an initial state $i$ and a final state $f$ is
\begin{equation}
\mathbf{d}_{i\rightarrow f} = \braket{f|\mathbf{r}|i},
    \label{eq:DME}
\end{equation}
where $\mathbf{r}$ is the position operator, and the states $\ket{i}$ and $\ket{f}$ are obtained from the many-body calculations described above. 
\section{Results}
We now demonstrate our atomistic approach to the optical excitation dynamics in InAsP/InP nanowire QDs.
First, tight-binding and configuration-interaction calculations are used to determine the single- and multi-exciton states of the system. 
From these states we evaluate the dipole matrix elements and spontaneous emission rates, characterizing the radiative properties of excitonic complexes. 
Next, we compute the phonon spectral density, providing a fully atomistic description of exciton–phonon coupling in this material system. 
Finally, we demonstrate the application of these quantities by calculating the emission brightness as a function of detuning and pulse strength, highlighting the sensitivity of this observable to the phonon spectral density.
\begin{figure*}[ht]    \includegraphics[width=\linewidth]{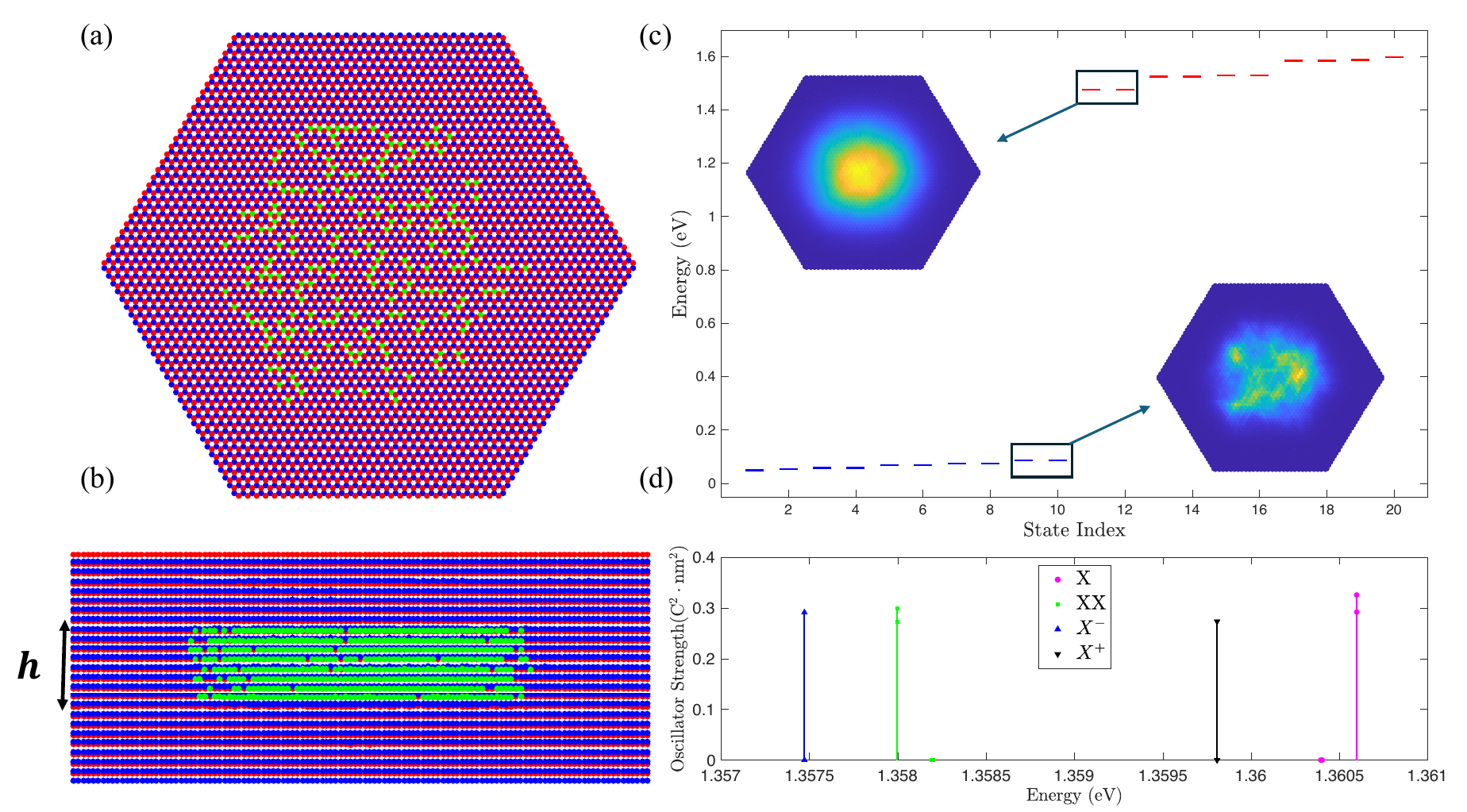}
    \caption{(a) Top of view of the InAsP/InP QD. The QD is formed in the center by replacement of Phosphorus  (blue circles)  by Arsenic (green circles) forming a cylindrical QD in the center. (b) Side view of the structure highlighting the spatial distribution of the active region. Panel (c) shows the single-particle TB spectrum with the corresponding charge densities for the top valence band state and lowest-energy conduction band state. (d) Low-energy exciton and multi-exciton spectral lines, showing the exciton neutral exciton ($X$), biexciton ($XX$), negative trion ($X^-$), and positive trion ($X^+$) transitions. 
} 
\label{fig:fig2}
 \end{figure*}
\subsection{Atomistic modeling of excitonic complexes}
We consider an InAsP/InP QD, a system whose electronic and optical properties are well established~\cite{cygorek202Atomistic}. The InAsP/InP QD is formed by taking a computational box consisting of a thick hexagonal slab with a diameter of about 24 nm and a width of about 8 nm of the InP host material. Then, in the center of the structure, in a hexagonal area of width of about 2.5 nm and a diameter of about 14.5 nm, 20$\%$ of the Phosphorus atoms are replaced with Arsenic. This center region forms a QD with a geometry similar to those in real devices~\cite{Laferriere2021}. Fig.~\ref{fig:fig2}(a) shows a top-down view of the QD structure, while Fig.~\ref{fig:fig2}(b) shows a side view. The structure as a whole contains about 130,000 atoms. We then solve the TB Hamiltonian in Eq.~\ref{eq: TBHam} to obtain the single-particle spectra and wavefunctions shown in Fig.~\ref{fig:fig2}(c). The band gap is about 1.4 eV. The states come in doublets corresponding to Kramers pairs. We show the charge density of the top of the VB state  (1s hole) and the bottom of the CB state (1s electron).  The electron wave function is highly symmetric, whereas the hole wave function is more sensitive to the randomness of the Arsenic distribution inside the QD. The latter thus deviates from the idealized Gaussian wave function form assumed in analytic treatments.

We then perform many-body calculations for $X$, $XX$, $X^+$, and $X^-$, 
by diagonalizing the Hamiltonian in Eq.~\eqref{eq:MBHam} within the corresponding configuration space defined by Eq.~\eqref{MB:WF}, taking into account 8 electron and 8 hole states. Here, the basis comprises single electron–hole pair excitations for excitons, charged single-pair excitations for trions, and two-pair excitations for biexcitons. For the charge neutral exciton, there are four low-energy states. The lowest two states are dark, and the two at slightly higher energy are bright. We denote the two bright excitons as $X_1$, and $X_2$.  We note that there is a fine structure splitting between the two bright excitons, but due to the high symmetry of the dot, this splitting is very small. The lowest-energy trion states of interest are a Kramers degenerate doublet, while the lowest-energy biexciton state of interest is nondegenerate. Now, with the correlated many-body wavefunctions of these excitonic complexes, we can proceed to bridge the gap between atomistic simulations and open quantum systems by computing the spontaneous emission rate, and the phonon spectral density.

\subsection{Dipole matrix elements and spontaneous emission rates}
We now proceed to compute the dipole matrix elements defined in Eq.~\eqref{eq:DME}. 
This requires an initial and final state which were obtained in the previous section. When the initial state is the charge neutral exciton $X_1$, $X_2$, the final state corresponds to the charge neutral ground state $GS$. For $X^-$, the final state is the charged  ground state containing one extra electron denoted by $e^-$. For $X^+$ the final state is the positively charged ground state $h^+$, and for the biexciton $XX$, it is an $X$ state. 

We then compute the corresponding spectral lines. 
Figure~\ref{fig:fig2}(d) displays the lowest exciton states ($X$), the lowest biexciton state ($XX$), and the lowest Kramers-degenerate trion states ($X^-$, $X^+$). 
The line heights correspond to the squared dipole matrix elements for the respective transitions. 
We observe dipole strengths that are relatively similar in magnitude.
\begin{table}[t]
\centering
\caption{Spontaneous emission rates for the lowest-energy excitonic complexes.}
\begin{tabular}{lc}
\hline\hline
Transition & Lifetime (ps) \\
\hline
$X_1\rightarrow GS$   &  927.8\\
$X_2\rightarrow GS$   &  1035\\
$XX \rightarrow X_1$  &  1017\\
$XX \rightarrow X_2$  &  1113\\
$X^-\rightarrow e^-$ &  1045\\
$X^+\rightarrow h^+$ &  1107\\
\hline\hline
\label{tb:tb1}
\end{tabular}
\label{tab:emission_rates}
\end{table}

Now that the transition dipole elements for the excitonic states of interest are available, we can compute the spontaneous emission rates of the optically active complexes using Eq.~\eqref{eq:SERate}. The refractive index $n(\omega)$ in the long wave-length limit is taken as 3.46. The resulting lifetimes, obtained as the inverse of the spontaneous emission rates, are summarized in Table~\ref{tb:tb1}. Note that we find variations of up to 10\% in the lifetimes occur for different random alloying. For InAsP/InP QDs of comparable dimensions, exciton lifetimes of about 1 ns at cryogenic temperatures are reported from experiments~\cite{hostein2008timeLifeTimeExperiment}, in good agreement with our atomistic theory.  
\begin{figure}[ht]    \includegraphics[width=\linewidth]{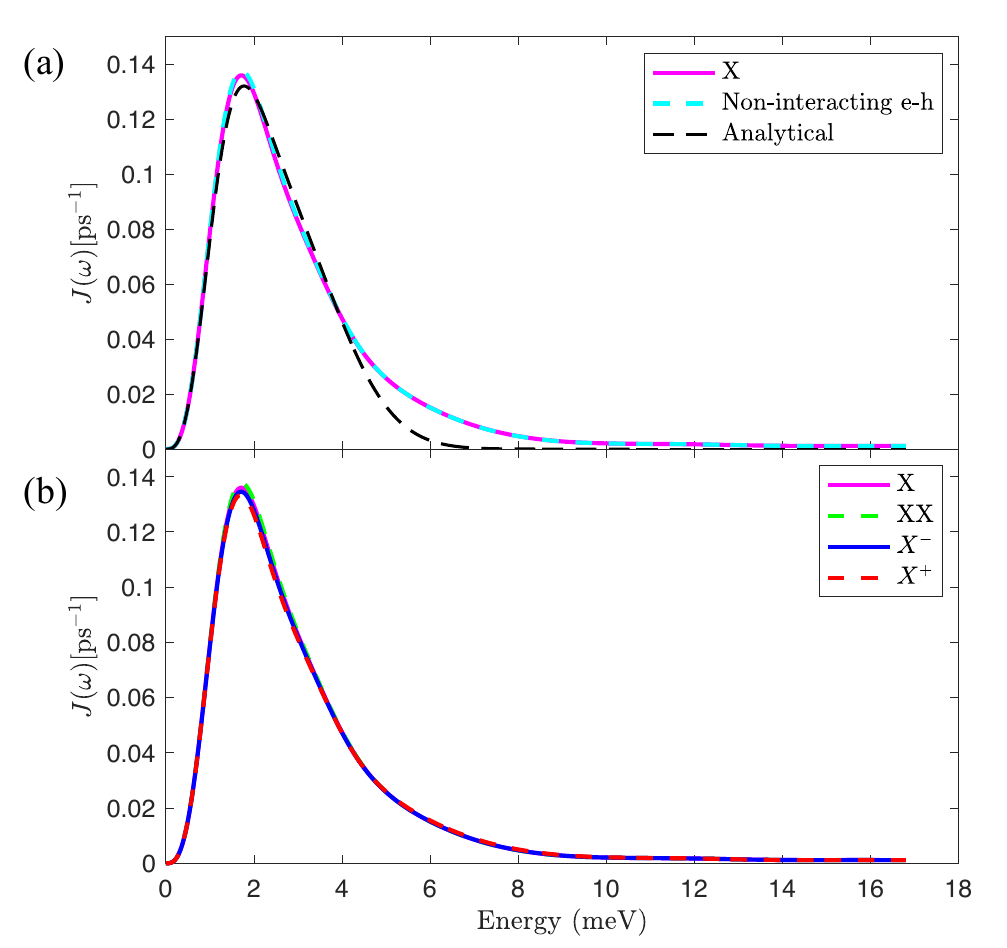}
    \caption{(a) Comparison of the numerically obtained phonon spectral density for the neutral exciton, with the non-correlated electron-hole only contribution and the corresponding analytical expression defined in Eq.~\eqref{eq: AnalyticalSpectralDensity}.
(b) Phonon spectral densities for the lowest-energy excitonic complexes, including the exciton ($X$), biexciton ($XX$), and trions ($X^-$, $X^+$), showing nearly identical behavior across all cases. 
}
\label{fig:fig3}
\end{figure}
\begin{figure}[ht]
    \includegraphics[width=\linewidth]{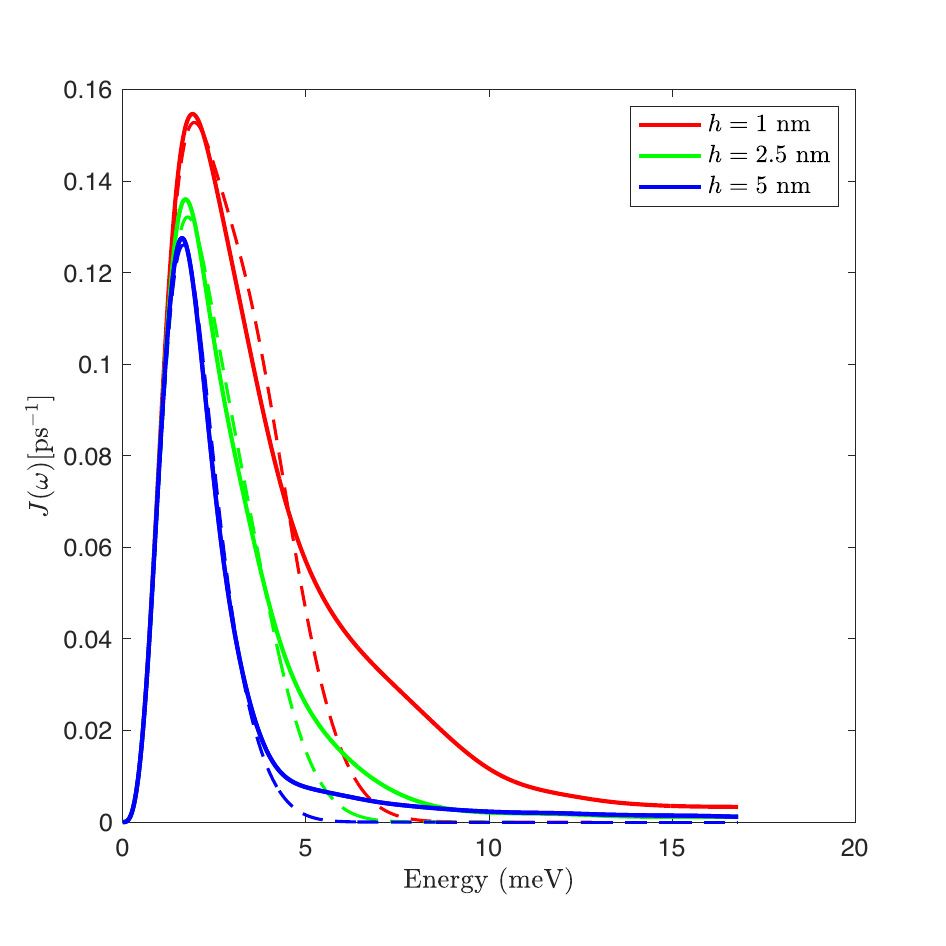}
    \caption{Phonon spectral densities for different QD heights $h$. Solid lines show the numerically computed spectral densities, while dashed lines indicate the corresponding analytical results. 
    }
    \label{fig:fig4}
\end{figure}
 \begin{figure*}[ht]
    \includegraphics[width=\linewidth]{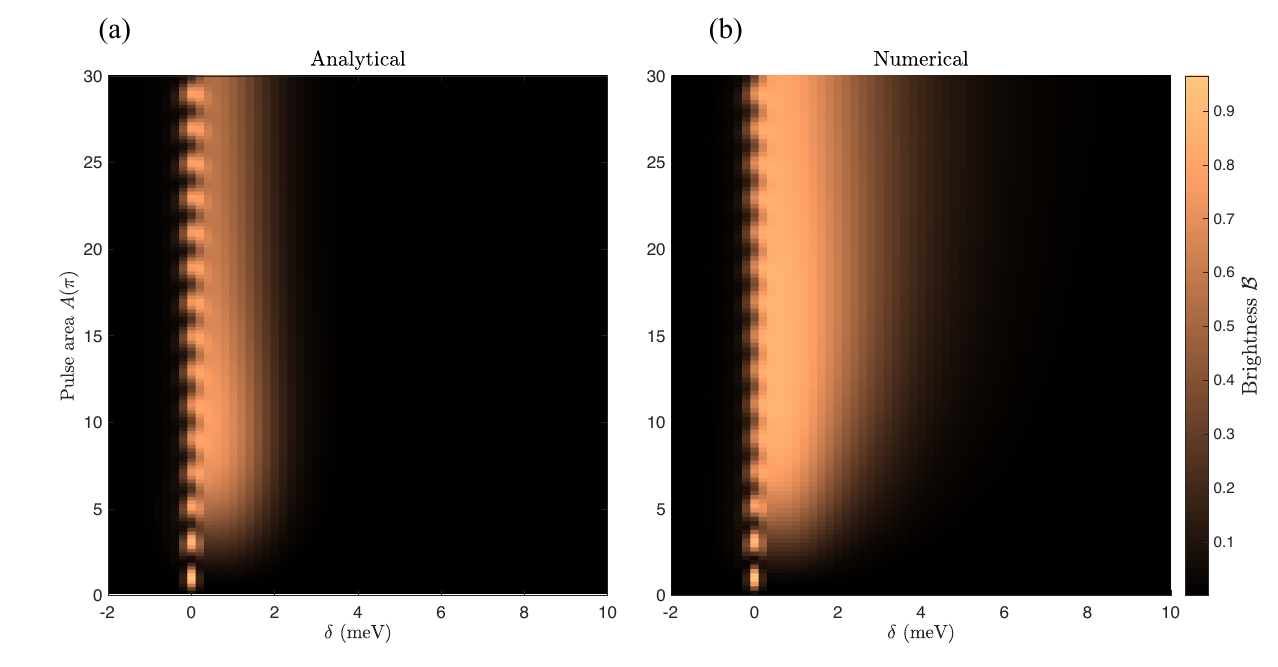}
    \caption{Calculated brightness $\mathcal{B}$ of the InAsP/InP QD 
    as a function of pulse area $A$ and detuning $\delta$. 
    The left panel (a) shows results obtained using the analytical phonon spectral density, 
    while the right panel (b) displays results based on the numerically computed spectral density. 
    The comparison highlights the impact of atomistic corrections to the exciton–phonon coupling on an experimentally relevant observable.
    }
    \label{fig:fig5}
\end{figure*}
\subsection{Phonon Spectral Density}
We now turn to the final ingredient required for the full simulation: the coupling of the QD to phonons, evaluated for each of the four excitonic complexes $X$, $XX$, $X^+$, and $X^-$. 
Owing to the high symmetry of the dot, it is sufficient to consider only one of the two low-energy bright exciton states, and, due to Kramers degeneracy, a single trion state. 
To compute the phonon spectral density defined in Eq.~\eqref{eq:PhononSpectralDensity}, a reference state must be specified. For the exciton state $X$ this is simply ground states $GS$, while for trions the reference states are single electron and single hole states, respectively. 

In Fig.~\ref{fig:fig3} we compare the phonon spectral density of the neutral exciton obtained from our atomistic approach with and without configuration mixing, together with the commonly used analytical form defined in Eq.~\eqref{eq: AnalyticalSpectralDensity}.
The analytical form has material parameters that are fixed, but also contain the  parameters  $a_e$ and $a_h$ which correspond to electron and hole confinement lengths. These will depend on the QD geometry~\cite{Reiter2017}. $a_e$ and $a_h$ are adjusted to best reproduce the atomistic result. 
This fitting procedure yields effective confinement lengths $a_e=4.55$~nm and $a_h=2.1$~nm that provide the closest agreement between the atomistic calculation and the analytical model. 

The curve labeled ``X'' in Fig.~\ref{fig:fig3}(a) denotes the fully correlated exciton including interactions between configurations, while 
``non-interacting e-h'' corresponds to restricting the calculation to the lowest-energy electron–hole configuration and neglecting many-body correlations. 
We emphasize that the difference between the correlated and uncorrelated exciton curves is only minor, demonstrating that configuration mixing has little effect on the phonon spectral density in this QD. A similar trend is found for trions and biexcitons, even though these complexes are more strongly correlated. 
This behavior can be traced to the relatively large dielectric constant of III–V semiconductors (12.9 in our calculations), together with the small dot size. 
As a result, the single-particle level spacing dominates over the Coulomb interaction strength. We emphasize that this does not diminish the role of configuration interaction in the theoretical framework. In addition to determining binding energies, fine-structure splittings, and transition energies, configuration interaction is essential for establishing that correlation effects on the phonon spectral density are indeed small in the present regime, which is not necessarily known apriori.

When comparing the analytical form of Eq.~\eqref{eq: AnalyticalSpectralDensity} with our numerically computed phonon spectral density, we find deviations at higher energies, manifested as a broader tail. 
This behavior is consistent with $\mathbf{k}\cdot\mathbf{p}$ calculations by Stock \textit{et al.}~\cite{stock2011acoustickdotp} for cylindrical InGaAs/GaAs dots, where deviations from the analytical model were also observed. 
Note that both, our results and Ref.~\cite{stock2011acoustickdotp}, differ from findings based on analytical wave functions of Ref.~\cite{Reiter2017}, where no such tail is predicted even for non-spherical QDs. We attribute this to the fact that Ref.~\cite{Reiter2017} specifically considers Gaussian wave functions in three spatial directions, while the geometries considered here suggest a more hard-core-like confinement along the QD height direction. Note that this effect depends on the overall geometry rather than atomistic details.

At low energies, both, atomistic results with and without many-body correlations, are in excellent agreement with the analytical expression used frequently in literature, while at higher energies there is a broadening of the tail of the spectral density. The small difference between the correlated and uncorrelated exciton curves demonstrates that configuration mixing has only a minor effect on the phonon spectral density, thereby justifying the frequent approximation of neglecting excitonic correlations in this context.

We further analyze the phonon spectral density as a function of the quantum dot height $h$. 
Fig~\ref{fig:fig4} shows the calculated phonon spectral density for different confinement heights. 
From fits to the numerical data, we extract electron and hole confinement lengths of 
$a_e = 4.35\,\mathrm{nm}$ and $a_h = 1.85\,\mathrm{nm}$ for $h = 1\,\mathrm{nm}$, 
while for $h = 5\,\mathrm{nm}$ we obtain $a_e = 4.35\,\mathrm{nm}$ and $a_h = 2.55\,\mathrm{nm}$. 
For small dot heights, the strong anisotropy of the carrier wave functions results in a highly non-spherical confinement potential, leading to noticeable deviations between the numerically computed spectral density and the analytical expression based on a spherical approximation. 
As the height $h$ is increased, the vertical confinement becomes comparable to the lateral confinement length scale, and the effective carrier localization approaches a more isotropic, quasi-spherical geometry. 
In this regime, exemplified by $h = 5\,\mathrm{nm}$, the numerical spectral density converges toward the analytical result, with improved agreement in both the peak position and overall magnitude. 
This behavior highlights the strong geometric dependence of the phonon spectral density, particularly at higher frequencies.


\subsection{Simulation of the QD as an Open Quantum System}
An important application of QDs is their use as single-photon sources. The properties of the emitted photons depend on the the properties of the QD as well as on the external laser driving. In this context, phonon-assisted state preparation may be employed as it is easy to implement yet yield excellent single-photon properties, where the emitted photons are by design spectrally separated from the excitation laser. All that is required is off-resonant excitation with laser pulse, typically about 1-2 meV above the QD resonance~\cite{PRLCosacchiBrightness,ThomasPhononAssisted2021}.

Having determined all atomistic ingredients required to describe carrier–phonon and light–matter interactions, 
we are now in a position to investigate the performance of phonon-assisted state preparation of a QD from its atomistic structure. 
To this end, we feed the calculated phonon spectral density as well as the radiative decay rate into the simulation code ACE~\cite{ACE_code}, which simulates the dynamics of open quantum systems in a numerically exact fashion, i.e. to all orders in the QD-phonon interaction.

In the simulations we model the QD as an effective two-level system consisting of the ground state $\ket{GS}$ and a single bright exciton state $\ket{X_1}$. 
The system is driven by a Gaussian laser pulse that couples $\ket{GS}$ and $\ket{X_1}$ via the dipole interaction, while phonon coupling is included through the spectral density introduced above. 
The dynamics are thus governed by the driven two-level Hamiltonian in Eq.~\eqref{eq:Driven2lsysH}, with losses entering via the Lindblad operator in Eq.~\eqref{eq: LindbladOp}. 
The spontaneous emission rate is taken from Tab.~\ref{tab:emission_rates} for $X_1$. The environment Hamiltonian is defined in Eq.~\eqref{eq: GeneralizedBose} which is characterized by the phonon spectral densities computed for $X$ in the previous section. The explicit form of the Gaussian pulse is given as
\begin{equation}
    \Omega(t) = \frac{A}{\sqrt{2\pi}\sigma}e^{-\frac{1}{2}(t-t_c)^2/\sigma^2},
\end{equation}
where, $A$ determines the driving strength, $t_c$ is the pulse center and $\sigma =\tau_{FWHM}/(2\sqrt{2\ln{2}})$ with $\tau_{FWHM}$ being the full width at half maximum of the pulse. 

The emission brightness is computed as~\cite{PRLCosacchiBrightness} 
\begin{equation}
    \mathcal{B} = \gamma_X \int_{-\infty}^{\infty} dt \, \langle  a^\dagger a \rangle(t),
\end{equation}
where $\gamma_X$ is the radiative decay rate of the exciton, and 
$a^\dagger$ ($a$) denote photon creation (annihilation) operators. 

We evaluate the brightness at $T = 4$ K for a pulse width of $\tau_{\mathrm{FWHM}} = 10$ ps. 
Figure~\ref{fig:fig5} shows the calculated brightness as a function of pulse area $A$ and detuning $\delta$. 
The left panel, Fig.~\ref{fig:fig5}(a), uses the analytical phonon spectral density of Eq.~\eqref{eq: AnalyticalSpectralDensity}, while the right panel, Fig.~\ref{fig:fig5}(b), employs the atomistically computed spectral density. 
In the absence of phonons, regions of high brightness are confined to a narrow range around resonance~\cite{PRLCosacchiBrightness}. 
Including phonons enables phonon-assisted transitions, where detunings are compensated by the emission of LA phonons, resulting in finite brightness away from resonance. 
The broader tail of the atomistic spectral density [Fig.~\ref{fig:fig3}(a)] directly manifests in Fig.~\ref{fig:fig5}(b) as a substantial broadening of the region of high brightness. 
We observe that the emission brightness $\mathcal{B}$ at fixed pulse area $A$ and detuning $\delta$ can differ substantially between the atomistically computed and analytical phonon spectral densities. As a representative example, for $A = 12\pi$ and $\delta = 3$~meV, the numerical spectral density yields a brightness of $\mathcal{B} \approx 0.29$, whereas the analytical model predicts a substantially smaller value of $\mathcal{B} \approx 0.034$. This corresponds to nearly an order-of-magnitude enhancement in the predicted brightness when using the atomistic spectral density.

\section{Discussion}
In this work, we have developed a fully atomistic framework for computing exciton–phonon coupling in semiconductor quantum dots that directly bridges microscopic electronic-structure calculations with non-Markovian open-quantum-system dynamics. Starting from an ab initio–parametrized tight-binding description, we combined configuration-interaction methods with a microscopic evaluation of electron–phonon coupling matrix elements, thereby obtaining phonon spectral densities without relying on phenomenological confinement models. This approach enables a nearly parameter-free description of carrier–phonon interactions that incorporates realistic dot geometries, atomistic wave functions, and many-body correlations.

Applying this framework to an InAsP/InP nanowire quantum dot, we obtain several insights:
First, fitting an established analytical form of a spectral density to the one obtained from atomistic simulations, we find that the low-frequency part is matched very well. This is especially important as the low-frequency part of $J(\omega)$ affects the qualitative behavior, e.g., of the free decay of coherences, and it determines the quantum dissipative phase of the corresponding spin-boson model~\cite{Leggett}. It affects in particular the stability of collective emission from multiple QDs~\cite{Wiercinski2023}. 
However, the atomistically derived spectral density generally shows a high-energy tail that is not reproduced by a fitted analytical form. This feature is shown to have a sizable impact on phonon-assisted transitions for large detunings. While explicitly explored this on the example of phonon-assisted state preparation~\cite{PRLCosacchiBrightness}, this finding is expected to generalize to other applications involving phonon-assisted transitions. For example, in the biexciton-exciton cascade for antibinding biexcitons~\cite{antibindingBX} or in the context of cavity feeding in solid-state cavity-QED devices~\cite{CavityFeeding_Kaer, CavityFeeding_Cygorek} high-energy features in the phonon spectral density make it challenging to prevent unwanted occupation of low-lying states. Similarly, high-energy tails affect relaxation of orbital hole qubits~\cite{orbitalhole} and increase the required excitation power to achieve the regime of dynamical decoupling~\cite{ReappearanceVagov,ReappearanceHanschke}. As we find that the high-energy features reduce when height and width of the QD become similar, more isotropic geometries are advantageous to avoid detrimental phonon-assisted transition. However, large aspect ratios can be favorable if off-resonant phonon-assisted state preparation is desired.

Another question addressed with microscopic simulations is how many-body correlations in excitonic complexes influence the coupling to phonons. While excitonic wave functions are generally not simple products of single-particle states, we find that the mixing of configurations has no visible effect on the effective coupling to the phonons. Moreover, phonon spectral densities for charged excitons (trions) are also virtually indistinguishable from those of neutral excitons in the same structure. This, in hindsight, justifies the general practice in the field, where excitonic correlations are largely ignored in the discussion of phonon effects.

Looking ahead, the presented framework opens several promising directions. It provides a natural route toward geometry-by-design optimization of quantum dots, where growth parameters can be tailored to engineer phonon spectral densities favorably. Second, by revisiting approximations made when computing the exciton-phonon coupling, and given the appropriate material paramaters the approach is readily extendable to emerging low-dimensional material platforms, such as quantum dots in transition-metal dichalcogenides, van der Waals heterostructures, and Moir\'{e}-defined quantum emitters, where reduced dimensionality and strong confinement are expected.

\appendix
\section{Calculation of Form Factors}
\label{app:FF}
Here we describe how we compute the form factors $\varrho_{n}(\textbf{k})$ given in Eq.~\eqref{eq:FFe} and \eqref{eq:FFh}. The form factors are defined as
\begin{equation}
    \varrho_{n}(\textbf{k}) = \int d\textbf{r}e^{i\textbf{k}\cdot\textbf{r}}\psi_{n}^{*}\left(\textbf{r}\right)\psi_{n}\left(\textbf{r}\right).
    \label{app: FormFactorDef}
\end{equation}
Now plugging in our TB wavefunctions defined in Eq.~\eqref{eq: TBWF} we get
\begin{equation}
\begin{split}
   \varrho_{n}(\textbf{k}) = &\sum_{\alpha,i,\alpha',i'}C^{n}_{\alpha',i'}(C^{n}_{\alpha,i})^* \\
   \times&\int d\textbf{r}e^{i\textbf{k}\cdot\textbf{r}}\phi_{\alpha}^{*}\left(\textbf{r}-\textbf{R}^0_i\right)\phi_{\alpha'}\left(\textbf{r}-\textbf{R}^0_{i'}\right)
    \end{split}
\end{equation}
To reduce the numerical complexity, and due to the localized and orthogonal nature of the atomic orbitals 
it is sufficient to consider $i=i'$ and  $\alpha=\alpha'$. We then approximate the density as $\phi_{\alpha}^{*}\left(\textbf{r}-\textbf{R}^0_i\right)\phi_{\alpha}\left(\textbf{r}-\textbf{R}^0_{i}\right)\approx\delta(\textbf{r}-\textbf{R}^0_{i})$ This gives 
\begin{equation}
    \hat{\varrho}_n(\textbf{k}) = \sum_{\alpha,i}|C^{n}_{\alpha,i}|^2e^{i\textbf{k}\cdot\textbf{R}_i}.
\end{equation}
We note that if one assumes a spherically symmetric parabolic confining potential for both conduction band and valence band (e/h), then
\begin{equation}
\psi^{par}_{e/h}(\textbf{r})=\left(d_{e/h}\sqrt{\pi}\right)^{-\frac{3}{2}}e^{-\frac{r^2}{2d^2_{e/h}}}    
\end{equation}
and the integral in Eq.~\eqref{app: FormFactorDef} is solved easily giving the familiar Gaussian form \cite{Krummheuer2002PRB,Nazir_2016,Vagov2002PRB}
\begin{equation}
\varrho^{par}_{e/h}(\textbf{r})=e^{-\frac{d^2_{e/h}|\textbf{k}|^2}{4}}.    
\end{equation}

\begin{acknowledgments}
Y.~S. and M.~C. acknowledge funding by the Return Program of the State of North Rhine-Westphalia. Y.~S. acknowledges financial support by the Deutsche Forschungsgemeinschaft (DFG, German Research Foundation) through the Würzburg-Dresden Cluster of Excellence ctd.qmat – Complexity, Topology and Dynamics in Quantum Matter (EXC 2147, project-id 390858490).
\end{acknowledgments}
\nocite{apsrev42Control}
\bibliographystyle{apsrev4-2}
\bibliography{Bibliography}

@CONTROL{apsrev42Control,
  author = "08",
  editor = "1",
  pages = "0",
  title = "0",
  year = "1"
}

@article{cygorek202Atomistic,
  title = {Atomistic theory of electronic and optical properties of InAsP/InP nanowire quantum dots},
  author = {Cygorek, Moritz and Korkusinski, Marek and Hawrylak, Pawel},
  journal = {Phys. Rev. B},
  volume = {101},
  issue = {7},
  pages = {075307},
  numpages = {13},
  year = {2020},
  month = {Feb},
  publisher = {American Physical Society},
  doi = {10.1103/PhysRevB.101.075307}
}

@article{CIPSI,
  title = {Accurate and efficient description of interacting carriers in quantum nanostructures by selected configuration interaction and perturbation theory},
  author = {Cygorek, Moritz and Otten, Matthew and Korkusinski, Marek and Hawrylak, Pawel},
  journal = {Phys. Rev. B},
  volume = {101},
  issue = {20},
  pages = {205308},
  numpages = {10},
  year = {2020},
  month = {May},
  publisher = {American Physical Society},
  doi = {10.1103/PhysRevB.101.205308}
}

@article{DnC,
  title = {Sublinear Scaling in Non-Markovian Open Quantum Systems Simulations},
  volume = {14},
  ISSN = {2160-3308},
  DOI = {10.1103/physrevx.14.011010},
  number = {1},
  pages = {011010},
  journal = {Phys. Rev. X},
  publisher = {American Physical Society (APS)},
  author = {Cygorek,  Moritz and Keeling,  Jonathan and Lovett,  Brendon W. and Gauger,  Erik M.},
  year = {2024},
  month = feb 
}

@article{ACE,
  title = {Simulation of open quantum systems by automated compression of arbitrary environments},
  volume = {18},
  ISSN = {1745-2481},
  DOI = {10.1038/s41567-022-01544-9},
  number = {6},
  journal = {Nat. Phys.},
  publisher = {Springer Science and Business Media LLC},
  author = {Cygorek,  Moritz and Cosacchi,  Michael and Vagov,  Alexei and Axt,  Vollrath Martin and Lovett,  Brendon W. and Keeling,  Jonathan and Gauger,  Erik M.},
  year = {2022},
  month = mar,
  pages = {662–668}
}

@article{ACE_code,
  title = {ACE: A general-purpose non-Markovian open quantum systems simulation toolkit based on process tensors},
  volume = {161},
  ISSN = {1089-7690},
  DOI = {10.1063/5.0221182},
  pages = {074111},
  number = {7},
  journal = {J. Chem. Phys.},
  publisher = {AIP Publishing},
  author = {Cygorek,  Moritz and Gauger,  Erik M.},
  year = {2024},
  month = aug 
}

@article{Laferriere2021,
  title = {Systematic study of the emission spectra of nanowire quantum dots},
  volume = {118},
  ISSN = {1077-3118},
  DOI = {10.1063/5.0045880},
  pages = {161107},
  number = {16},
  journal = {Appl. Phys. Lett.},
  publisher = {AIP Publishing},
  author = {Laferrière,  Patrick and Yeung,  Edith and Korkusinski,  Marek and Poole,  Philip J. and Williams,  Robin L. and Dalacu,  Dan and Manalo,  Jacob and Cygorek,  Moritz and Altintas,  Abdulmenaf and Hawrylak,  Pawel},
  year = {2021},
  month = apr 
}

@article{Krummheuer2002PRB,
  title = {Theory of pure dephasing and the resulting absorption line shape in semiconductor quantum dots},
  author = {Krummheuer, B. and Axt, V. M. and Kuhn, T.},
  journal = {Phys. Rev. B},
  volume = {65},
  issue = {19},
  pages = {195313},
  numpages = {12},
  year = {2002},
  month = {May},
  publisher = {American Physical Society},
  doi = {10.1103/PhysRevB.65.195313}
}

@article{Krummheuer2005PRB,
  title = {Pure dephasing and phonon dynamics in GaAs- and GaN-based quantum dot structures: Interplay between material parameters and geometry},
  volume = {71},
  ISSN = {1550-235X},
  DOI = {10.1103/physrevb.71.235329},
  pages = {235329},
  number = {23},
  journal = {Phys. Rev. B},
  publisher = {American Physical Society (APS)},
  author = {Krummheuer,  B. and Axt,  V. M. and Kuhn,  T. and D’Amico,  I. and Rossi,  F.},
  year = {2005},
  month = jun 
}

@article{Krzykowski2020,
  title = {Hole spin-flip transitions in a self-assembled quantum dot},
  volume = {102},
  ISSN = {2469-9969},
  DOI = {10.1103/physrevb.102.205301},
  pages = {205301},
  number = {20},
  journal = {Phys. Rev. B},
  publisher = {American Physical Society (APS)},
  author = {Krzykowski,  Mateusz and Gawarecki,  Krzysztof and Machnikowski,  Paweł},
  year = {2020},
  month = nov 
}

@article{Nazir_2016,
	author = {Nazir, Ahsan and McCutcheon, Dara P S},
	doi = {10.1088/0953-8984/28/10/103002},
	journal = {J. Phys.: Condens. Matter},
	month = {feb},
	number = {10},
	pages = {103002},
	publisher = {IOP Publishing},
	title = {Modelling exciton--phonon interactions in optically driven quantum dots},
	volume = {28},
	year = {2016},
}

@article{hostein2008timeLifeTimeExperiment,
  title={Time-resolved characterization of InAsP/ InP quantum dots emitting in the C-band telecommunication window},
  author={Hostein, Richard and Michon, Adrien and Beaudoin, Gregoire and Gogneau, Noelle and Patriache, Gilles and Marzin, J-Y and Robert-Philip, Isabelle and Sagnes, Isabelle and Beveratos, Alexios},
  journal={Appl. Phys. Lett.},
  pages = {073106},
  volume={93},
  number={7},
  year={2008},
  publisher={AIP Publishing}
}

@article{stock2011acoustickdotp,
  title = {Acoustic and optical phonon scattering in a single In(Ga)As quantum dot},
  author = {Stock, Erik and Dachner, Matthias-Rene and Warming, Till and Schliwa, Andrei and Lochmann, Anatol and Hoffmann, Axel and Toropov, Aleksandr I. and Bakarov, Askhat K. and Derebezov, Ilya A. and Richter, Marten and others},
  journal = {Phys. Rev. B},
  volume = {83},
  issue = {4},
  pages = {041304(R)},
  numpages = {4},
  year = {2011},
  month = {Jan},
  publisher = {American Physical Society},
  doi = {10.1103/PhysRevB.83.041304}
}

@article{PRLCosacchiBrightness,
  title = {Emission-Frequency Separated High Quality Single-Photon Sources Enabled by Phonons},
  author = {Cosacchi, M. and Ungar, F. and Cygorek, M. and Vagov, A. and Axt, V. M.},
  journal = {Phys. Rev. Lett.},
  volume = {123},
  issue = {1},
  pages = {017403},
  numpages = {5},
  year = {2019},
  month = {Jul},
  publisher = {American Physical Society},
  doi = {10.1103/PhysRevLett.123.017403},
}

@misc{floquet,
      title={Robust entangled photon generation enabled by single-shot Floquet driving}, 
      author={Jun-Yong Yan and Paul C. A. Hagen and Hans-Georg Babin and Wei E. I. Sha and Andreas D. Wieck and Arne Ludwig and Chao-Yuan Jin and Vollrath M. Axt and Da-Wei Wang and Moritz Cygorek and Feng Liu},
      year={2025},
      eprint={2504.02753},
      archivePrefix={arXiv},
      primaryClass={quant-ph}
}

@article{SUPER,
  title = {Swing-Up of Quantum Emitter Population Using Detuned Pulses},
  author = {Bracht, Thomas K. and Cosacchi, Michael and Seidelmann, Tim and Cygorek, Moritz and Vagov, Alexei and Axt, V. Martin and Heindel, Tobias and Reiter, Doris E.},
  journal = {PRX Quantum},
  volume = {2},
  issue = {4},
  pages = {040354},
  numpages = {11},
  year = {2021},
  month = {Dec},
  publisher = {American Physical Society},
  doi = {10.1103/PRXQuantum.2.040354}
}

@article{Heinisch2024,
  title = {Swing-up dynamics in quantum emitter cavity systems: Near ideal single photons and entangled photon pairs},
  author = {Heinisch, Nils and K\"ocher, Nikolas and Bauch, David and Schumacher, Stefan},
  journal = {Phys. Rev. Res.},
  volume = {6},
  issue = {1},
  pages = {L012017},
  numpages = {7},
  year = {2024},
  month = {Jan},
  publisher = {American Physical Society},
  doi = {10.1103/PhysRevResearch.6.L012017}
}

@article{BrachtOpticaQuantum2023,
author = {Thomas K. Bracht and Moritz Cygorek and Tim Seidelmann and Vollrath Martin Axt and Doris E. Reiter},
journal = {Optica Quantum},
number = {2},
pages = {103--107},
publisher = {Optica Publishing Group},
title = {Temperature-independent almost perfect photon entanglement from quantum dots via the SUPER scheme},
volume = {1},
month = {Dec},
year = {2023},
doi = {10.1364/OPTICAQ.498559}
}

@article{SeidelmannLimit,
  title = {Two-Photon Excitation Sets Limit to Entangled Photon Pair Generation from Quantum Emitters},
  author = {Seidelmann, T. and Schimpf, C. and Bracht, T. K. and Cosacchi, M. and Vagov, A. and Rastelli, A. and Reiter, D. E. and Axt, V. M.},
  journal = {Phys. Rev. Lett.},
  volume = {129},
  issue = {19},
  pages = {193604},
  numpages = {7},
  year = {2022},
  month = {Nov},
  publisher = {American Physical Society},
  doi = {10.1103/PhysRevLett.129.193604},
}

@article{SeidelmannStrongToWeak,
  title = {From strong to weak temperature dependence of the two-photon entanglement resulting from the biexciton cascade inside a cavity},
  author = {Seidelmann, T. and Ungar, F. and Cygorek, M. and Vagov, A. and Barth, A. M. and Kuhn, T. and Axt, V. M.},
  journal = {Phys. Rev. B},
  volume = {99},
  issue = {24},
  pages = {245301},
  numpages = {13},
  year = {2019},
  month = {Jun},
  publisher = {American Physical Society},
  doi = {10.1103/PhysRevB.99.245301}
}

@article{ARPSimon2011,
  title = {Robust Quantum Dot Exciton Generation via Adiabatic Passage with Frequency-Swept Optical Pulses},
  volume = {106},
  ISSN = {1079-7114},
  DOI = {10.1103/physrevlett.106.166801},
  pages = {166801},
  number = {16},
  journal = {Phys. Rev. Lett},
  publisher = {American Physical Society (APS)},
  author = {Simon,  C.-M. and Belhadj,  T. and Chatel,  B. and Amand,  T. and Renucci,  P. and Lemaitre,  A. and Krebs,  O. and Dalgarno,  P. A. and Warburton,  R. J. and Marie,  X. and Urbaszek,  B.},
  year = {2011},
  month = apr 
}

@article{ARPWu2011,
  title = {Population Inversion in a Single InGaAs Quantum Dot Using the Method of Adiabatic Rapid Passage},
  volume = {106},
  ISSN = {1079-7114},
  DOI = {10.1103/physrevlett.106.067401},
  pages = {067401},
  number = {6},
  journal = {    Phys. Rev. Lett.},
  publisher = {American Physical Society (APS)},
  author = {Wu,  Yanwen and Piper,  I. M. and Ediger,  M. and Brereton,  P. and Schmidgall,  E. R. and Eastham,  P. R. and Hugues,  M. and Hopkinson,  M. and Phillips,  R. T.},
  year = {2011},
  month = feb 
}

@article{MuljarovZimmermann,
  title = {Dephasing in Quantum Dots: Quadratic Coupling to Acoustic Phonons},
  author = {Muljarov, E. A. and Zimmermann, R.},
  journal = {Phys. Rev. Lett.},
  volume = {93},
  issue = {23},
  pages = {237401},
  numpages = {4},
  year = {2004},
  month = {Nov},
  publisher = {American Physical Society},
  doi = {10.1103/PhysRevLett.93.237401}
}

@article{Kaldewey2017,
  title = {Demonstrating the decoupling regime of the electron-phonon interaction in a quantum dot using chirped optical excitation},
  author = {Kaldewey, Timo and L\"uker, Sebastian and Kuhlmann, Andreas V. and Valentin, Sascha R. and Chauveau, Jean-Michel and Ludwig, Arne and Wieck, Andreas D. and Reiter, Doris E. and Kuhn, Tilmann and Warburton, Richard J.},
  journal = {Phys. Rev. B},
  volume = {95},
  issue = {24},
  pages = {241306},
  numpages = {5},
  year = {2017},
  month = {Jun},
  publisher = {American Physical Society},
  doi = {10.1103/PhysRevB.95.241306}
}

@article{FranzCarl2025,
	author = {M{\o}rch Nielsen, Carl Emil and Fischer, Franz and Bester, Gabriel},
	date = {2025/02/08},
	doi = {10.1038/s41699-025-00532-w},
	id = {M{\o}rch Nielsen2025},
	isbn = {2397-7132},
	journal = {npj 2D Mater. Appl.},
	number = {1},
	pages = {11},
	title = {Beyond the K-valley: exploring unique trion states in indirect band gap monolayer WSe2},
	volume = {9},
	year = {2025},
}

@article{zielinski2010atomistic,
  title={Atomistic tight-binding theory of multiexciton complexes in a self-assembled InAs quantum dot},
  author={Zieli{\'n}ski, Micha{\l} and Korkusi{\'n}ski, M and Hawrylak, P},
  journal={Phys. Rev. B},
  volume={81},
  number={8},
  pages={085301},
  year={2010},
  publisher={APS}
}

@article{jaskolski2006strain,
  title={Strain effects on the electronic structure of strongly coupled self-assembled InAs/ GaAs quantum dots: Tight-binding approach},
  author={Jask{\'o}lski, W and Zieli{\'n}ski, M and Bryant, Garnett W and Aizpurua, Javier},
  journal={Phys. Rev. B},
  volume={74},
  number={19},
  pages={195339},
  year={2006},
  publisher={APS}
}

@article{bester2003pseudopotential,
  title = {Pseudopotential calculation of the excitonic fine structure of million-atom self-assembled ${\mathrm{In}}_{1\ensuremath{-}x}{\mathrm{Ga}}_{x}\mathrm{A}\mathrm{s}/\mathrm{G}\mathrm{a}\mathrm{A}\mathrm{s}$ quantum dots},
  author = {Bester, Gabriel and Nair, Selvakumar and Zunger, Alex},
  journal = {Phys. Rev. B},
  volume = {67},
  issue = {16},
  pages = {161306},
  numpages = {4},
  year = {2003},
  month = {Apr},
  publisher = {American Physical Society},
  doi = {10.1103/PhysRevB.67.161306},
}

@article{dalacu2009selective,
  title={Selective-area vapour--liquid--solid growth of InP nanowires},
  author={Dalacu, Dan and Kam, Alicia and Austing, D Guy and Wu, Xiaohua and Lapointe, Jean and Aers, Geof C and Poole, Philip J},
  journal={Nanotechnology},
  volume={20},
  number={39},
  pages={395602},
  year={2009},
  publisher={IOP Publishing}
}

@article{dalacu2012ultraclean,
  title={Ultraclean emission from InAsP quantum dots in defect-free wurtzite InP nanowires},
  author={Dalacu, Dan and Mnaymneh, Khaled and Lapointe, Jean and Wu, Xiaohua and Poole, Philip J and Bulgarini, Gabriele and Zwiller, Val and Reimer, Michael E},
  journal={Nano Lett.},
  volume={12},
  number={11},
  pages={5919--5923},
  year={2012},
  publisher={ACS Publications}
}

@article{dalacu2021tailoring,
  title={Tailoring the geometry of bottom-up nanowires: Application to high efficiency single photon sources},
  author={Dalacu, Dan and Poole, Philip J and Williams, Robin L},
  journal={Nanomaterials},
  volume={11},
  number={5},
  pages={1201},
  year={2021},
  publisher={MDPI}
}

@article{schweickert2018demandsinglePhoton,
  title={On-demand generation of background-free single photons from a solid-state source},
  author={Schweickert, Lucas and J{\"o}ns, Klaus D and Zeuner, Katharina D and Covre da Silva, Saimon Filipe and Huang, Huiying and Lettner, Thomas and Reindl, Marcus and Zichi, Julien and Trotta, Rinaldo and Rastelli, Armando and others},
  journal={Appl. Phys. Lett.},
  pages = {093106},
  volume={112},
  number={9},
  year={2018},
  publisher={AIP Publishing}
}

@article{huber2017highly,
  title={Highly indistinguishable and strongly entangled photons from symmetric GaAs quantum dots},
  author={Huber, Daniel and Reindl, Marcus and Huo, Yongheng and Huang, Huiying and Wildmann, Johannes S and Schmidt, Oliver G and Rastelli, Armando and Trotta, Rinaldo},
  journal={Nat. Commun.},
  volume={8},
  number={1},
  pages={15506},
  year={2017},
  publisher={Nature Publishing Group UK London}
}

@article{bimberg2002ingaasLaser,
  title={InGaAs-GaAs quantum-dot lasers},
  author={Bimberg, D and Kirstaedter, N and Ledentsov, NN and Alferov, Zh I and Kop'Ev, PS and Ustinov, VM},
  journal={IEEE J. Sel. Top. Quantum Electron.},
  volume={3},
  number={2},
  pages={196--205},
  year={2002},
  publisher={IEEE}
}

@article{fafard1996redlaser,
  title={Red-emitting semiconductor quantum dot lasers},
  author={Fafard, S and Hinzer, K and Raymond, S and Dion, M and McCaffrey, J and Feng, Y and Charbonneau, S},
  journal={Science},
  volume={274},
  number={5291},
  pages={1350--1353},
  year={1996},
  publisher={American Association for the Advancement of Science}
}

@article{schwagmann2011chipSinglePhotoWavefuide,
  title={On-chip single photon emission from an integrated semiconductor quantum dot into a photonic crystal waveguide},
  author={Schwagmann, Andre and Kalliakos, Sokratis and Farrer, Ian and Griffiths, Jonathan P and Jones, Geb AC and Ritchie, David A and Shields, Andrew J},
  journal={Appl. Phys. Lett.},
  pages = {261108},
  volume={99},
  number={26},
  year={2011},
  publisher={AIP Publishing}
}

@article{huber2020filter,
  title={Filter-free single-photon quantum dot resonance fluorescence in an integrated cavity-waveguide device},
  author={Huber, Tobias and Davanco, Marcelo and M{\"u}ller, Markus and Shuai, Yichen and Gazzano, Olivier and Solomon, Glenn S},
  journal={Optica},
  volume={7},
  number={5},
  pages={380--385},
  year={2020},
  publisher={Optical Society of America}
}

@article{ota2011spontaneous,
  title={Spontaneous two-photon emission from a single quantum dot},
  author={Ota, Yasutomo and Iwamoto, Satoshi and Kumagai, Naoto and Arakawa, Yasuhiko},
  journal={    Phys. Rev. Lett.},
  volume={107},
  number={23},
  pages={233602},
  year={2011},
  publisher={APS}
}

@article{liu2024dynamic,
  title={Dynamic resonance fluorescence in solid-state cavity quantum electrodynamics},
  author={Liu, Shunfa and Gustin, Chris and Liu, Hanqing and Li, Xueshi and Yu, Ying and Ni, Haiqiao and Niu, Zhichuan and Hughes, Stephen and Wang, Xuehua and Liu, Jin},
  journal={Nat. Photonics},
  volume={18},
  number={4},
  pages={318--324},
  year={2024},
  publisher={Nature Publishing Group UK London}
}

@article{qian2018two,
  title={Two-photon Rabi splitting in a coupled system of a nanocavity and exciton complexes},
  author={Qian, Chenjiang and Wu, Shiyao and Song, Feilong and Peng, Kai and Xie, Xin and Yang, Jingnan and Xiao, Shan and Steer, Matthew J and Thayne, Iain G and Tang, Chengchun and others},
  journal={    Phys. Rev. Lett.},
  volume={120},
  number={21},
  pages={213901},
  year={2018},
  publisher={APS}
}

@article{schumacher2012cavity,
  title={Cavity-assisted emission of polarization-entangled photons from biexcitons in quantum dots with fine-structure splitting},
  author={Schumacher, Stefan and F{\"o}rstner, Jens and Zrenner, Artur and Florian, Matthias and Gies, Christopher and Gartner, Paul and Jahnke, Frank},
  journal={Opt. express},
  volume={20},
  number={5},
  pages={5335--5342},
  year={2012},
  publisher={Optical Society of America}
}

@article{del2011generationMicroCavity,
  title={Generation of a two-photon state from a quantum dot in a microcavity},
  author={Del Valle, E and Gonzalez--Tudela, A and Cancellieri, E and Laussy, FP and Tejedor, C},
  journal={New J. Phys.},
  volume={13},
  number={11},
  pages={113014},
  year={2011},
  publisher={IOP Publishing}
}

@article{liu2025quantumNAture,
  title={Quantum correlations of spontaneous two-photon emission from a quantum dot},
  author={Liu, Shunfa and Wang, Yangpeng and Saleem, Yasser and Li, Xueshi and Liu, Hanqing and Yang, Cheng-Ao and Yang, Jiawei and Ni, Haiqiao and Niu, Zhichuan and Meng, Yun and others},
  journal={Nature},
  volume={643},
  number={8074},
  pages={1234--1239},
  year={2025},
  publisher={Nature Publishing Group UK London}
}

@article{Takagahara1993PRL,
  title = {Electron-phonon interactions and excitonic dephasing in semiconductor nanocrystals},
  author = {Takagahara, T.},
  journal = {Phys. Rev. Lett.},
  volume = {71},
  issue = {21},
  pages = {3577--3580},
  numpages = {0},
  year = {1993},
  month = {Nov},
  publisher = {American Physical Society},
  doi = {10.1103/PhysRevLett.71.3577},
}

@article{klimeck2007atomisticII,
  title={Atomistic simulation of realistically sized nanodevices using NEMO 3-D—Part II: Applications},
  author={Klimeck, Gerhard and Ahmed, Shaikh Shahid and Kharche, Neerav and Korkusinski, Marek and Usman, Muhammad and Prada, Marta and Boykin, Timothy B},
  journal={IEEE Trans. Electron Devices},
  volume={54},
  number={9},
  pages={2090--2099},
  year={2007},
  publisher={IEEE}
}

@article{zunger1998electronic,
  title={Electronic-structure theory of semiconductor quantum dots},
  author={Zunger, Alex},
  journal={MRS bulletin},
  volume={23},
  number={2},
  pages={35--42},
  year={1998},
  publisher={Cambridge University Press}
}

@article{klimeck2007atomisticI,
  title={Atomistic simulation of realistically sized nanodevices using NEMO 3-D—Part I: Models and benchmarks},
  author={Klimeck, Gerhard and Ahmed, Shaikh Shahid and Bae, Hansang and Kharche, Neerav and Clark, Steve and Haley, Benjamin and Lee, Sunhee and Naumov, Maxim and Ryu, Hoon and Saied, Faisal and others},
  journal={IEEE Trans. Electron Devices},
  volume={54},
  number={9},
  pages={2079--2089},
  year={2007},
  publisher={IEEE}
}

@article{hawrylak1996electronic,
  title={Electronic structure and optical properties of self-assembled quantum dots},
  author={Hawrylak, Pawel and Wojs, Arkadiusz},
  journal={Semicond. Sci. Technol.},
  volume={11},
  number={11S},
  pages={1516},
  year={1996},
  publisher={IOP Publishing}
}

@article{Vagov2002PRB,
  title = {Electron-phonon dynamics in optically excited quantum dots: Exact solution for multiple ultrashort laser pulses},
  author = {Vagov, A. and Axt, V. M. and Kuhn, T.},
  journal = {Phys. Rev. B},
  volume = {66},
  issue = {16},
  pages = {165312},
  numpages = {15},
  year = {2002},
  month = {Oct},
  publisher = {American Physical Society},
  doi = {10.1103/PhysRevB.66.165312},
}

@article{NarvaezBesterZungerPRB2005,
  title = {Excitons, biexcitons, and trions in self-assembled $(\mathrm{In}\text{,}\mathrm{Ga})\mathrm{As}/\mathrm{Ga}\mathrm{As}$ quantum dots: Recombination energies, polarization, and radiative lifetimes versus dot height},
  author = {Narvaez, Gustavo A. and Bester, Gabriel and Zunger, Alex},
  journal = {Phys. Rev. B},
  volume = {72},
  issue = {24},
  pages = {245318},
  numpages = {10},
  year = {2005},
  month = {Dec},
  publisher = {American Physical Society},
  doi = {10.1103/PhysRevB.72.245318},
}

@book{Scully1997,
  title     = {Quantum Optics},
  author    = {Scully, M. O. and Zubairy, M. S.},
  year      = {1997},
  publisher = {Cambridge University Press},
  address   = {Cambridge},
}

@article{senellart2017highSinglePhoton,
  title={High-performance semiconductor quantum-dot single-photon sources},
  author={Senellart, Pascale and Solomon, Glenn and White, Andrew},
  journal={Nat. Nanotechnol.},
  volume={12},
  number={11},
  pages={1026--1039},
  year={2017},
  publisher={Nature Publishing Group UK London}
}

@article{ThomasPhononAssisted2021,
  title = {Bright Polarized Single-Photon Source Based on a Linear Dipole},
  volume = {126},
  ISSN = {1079-7114},
  DOI = {10.1103/physrevlett.126.233601},
  pages = {233601},
  number = {23},
  journal = {    Phys. Rev. Lett.},
  publisher = {American Physical Society (APS)},
  author = {Thomas, S. E. and Billard, M. and Coste, N. and Wein, S. C. and Priya and Ollivier, H. and Krebs, O. and Tazaïrt, L. and Harouri, A. and Lemaitre, A. and others},
  year = {2021},
  month = jun 
}

@article{huber2018semiconductorEntangledPhotonPairs,
  title={Semiconductor quantum dots as an ideal source of polarization-entangled photon pairs on-demand: a review},
  author={Huber, Daniel and Reindl, Marcus and Aberl, Johannes and Rastelli, Armando and Trotta, Rinaldo},
  journal={J. Opt.},
  volume={20},
  number={7},
  pages={073002},
  year={2018},
  publisher={IOP Publishing}
}

@article{schwartz2016deterministicClusterStates,
  title={Deterministic generation of a cluster state of entangled photons},
  author={Schwartz, Ido and Cogan, Dan and Schmidgall, Emma R and Don, Yaroslav and Gantz, Liron and Kenneth, Oded and Lindner, Netanel H and Gershoni, David},
  journal={Science},
  volume={354},
  number={6311},
  pages={434--437},
  year={2016},
  publisher={American Association for the Advancement of Science}
}

@article{Lindner2009PhotonCluster,
  title = {Proposal for Pulsed On-Demand Sources of Photonic Cluster State Strings},
  author = {Lindner, Netanel H. and Rudolph, Terry},
  journal = {Phys. Rev. Lett.},
  volume = {103},
  issue = {11},
  pages = {113602},
  numpages = {4},
  year = {2009},
  month = {Sep},
  publisher = {American Physical Society},
  doi = {10.1103/PhysRevLett.103.113602},
}

@article{arakawa2020progressSinglePhoton,
  title={Progress in quantum-dot single photon sources for quantum information technologies: A broad spectrum overview},
  author={Arakawa, Yasuhiko and Holmes, Mark J},
  journal={Appl. Phys. Rev.},
  DOI = {10.1063/5.0010193},
  pages = {021309},
  volume={7},
  number={2},
  year={2020},
  publisher={AIP Publishing}
}

@article{Arakawa1982SemiconductingLasers,
    author = {Arakawa, Y. and Sakaki, H.},
    title = {Multidimensional quantum well laser and temperature dependence of its threshold current},
    journal = {Appl. Phys. Lett.},
    volume = {40},
    number = {11},
    pages = {939-941},
    year = {1982},
    month = {06},
    issn = {0003-6951},
    doi = {10.1063/1.92959},
}

@article{Reiter2017,
  title = {Phonon impact on optical control schemes of quantum dots: Role of quantum dot geometry and symmetry},
  volume = {96},
  ISSN = {2469-9969},
  DOI = {10.1103/physrevb.96.245306},
  pages = {245306},
  number = {24},
  journal = {Phys. Rev. B},
  publisher = {American Physical Society (APS)},
  author = {L\"{u}ker,  S. and Kuhn,  T. and Reiter,  D. E.},
  year = {2017},
  month = dec 
}

@book{mahanbook,
  title={Many-particle physics},
  author={Mahan, Gerald D},
  year={2013},
  publisher={Springer Science \& Business Media}
}

@article{wigger2021resonance,
  title={Resonance-fluorescence spectral dynamics of an acoustically modulated quantum dot},
  author={Wigger, Daniel and Wei{\ss}, Matthias and Lienhart, Michelle and M{\"u}ller, Kai and Finley, Jonathan J and Kuhn, Tilmann and Krenner, Hubert J and Machnikowski, Pawe{\l}},
  journal={Phys. Rev. Research},
  volume={3},
  number={3},
  pages={033197},
  year={2021},
  publisher={APS}
}

@article{niehues2018strain,
  title={Strain control of exciton--phonon coupling in atomically thin semiconductors},
  author={Niehues, Iris and Schmidt, Robert and Druppel, Matthias and Marauhn, Philipp and Christiansen, Dominik and Selig, Malte and Bergh{\"a}user, Gunnar and Wigger, Daniel and Schneider, Robert and Braasch, Lisa and others},
  journal={Nano Lett.},
  volume={18},
  number={3},
  pages={1751--1757},
  year={2018},
  publisher={ACS Publications}
}

@article{Groll_2021,
	author = {Groll, Daniel and Hahn, Thilo and Machnikowski, Pawe{\l} and Wigger, Daniel and Kuhn, Tilmann},
	doi = {10.1088/2633-4356/abcbeb},
	journal = {Mater. Quantum Technol.},
	month = {mar},
	number = {1},
	pages = {015004},
	publisher = {IOP Publishing},
	title = {Controlling photoluminescence spectra of hBN color centers by selective phonon-assisted excitation: a theoretical proposal},
	volume = {1},
	year = {2021},
}

@article{WiggerLA2020,
	author = {Daniel Wigger and Vage Karakhanyan and Christian Schneider and Martin Kamp and Sven H\"{o}fling and Pawe{\l} Machnikowski and Tilmann Kuhn and Jacek Kasprzak},
	doi = {10.1364/OL.385602},
	journal = {Opt. Lett.},
	month = {Feb},
	number = {4},
	pages = {919--922},
	publisher = {Optica Publishing Group},
	title = {Acoustic phonon sideband dynamics during polaron formation in a single quantum dot},
	volume = {45},
	year = {2020},
}

@article{FSS_Takagahara,
  title = {Theory of exciton doublet structures and polarization relaxation in single quantum dots},
  author = {Takagahara, T.},
  journal = {Phys. Rev. B},
  volume = {62},
  issue = {24},
  pages = {16840},
  numpages = {0},
  year = {2000},
  month = {Dec},
  publisher = {American Physical Society},
  doi = {10.1103/PhysRevB.62.16840}
}

@article{Leggett,
  title = {Dynamics of the dissipative two-state system},
  author = {Leggett, A. J. and Chakravarty, S. and Dorsey, A. T. and Fisher, Matthew P. A. and Garg, Anupam and Zwerger, W.},
  journal = {Rev. Mod. Phys.},
  volume = {59},
  issue = {1},
  pages = {1--85},
  numpages = {0},
  year = {1987},
  month = {Jan},
  publisher = {American Physical Society},
  doi = {10.1103/RevModPhys.59.1}
}

@article{antibindingBX,
  title = {Biexciton initialization by two-photon excitation in site-controlled quantum dots: The complexity of the antibinding state case},
  volume = {117},
  ISSN = {1077-3118},
  DOI = {10.1063/5.0011383},
  pages = {134001},
  number = {13},
  journal = {Appl. Phys. Lett.},
  publisher = {AIP Publishing},
  author = {Juska,  Gediminas and Ranjbar Jahromi,  Iman and Mattana,  Francesco and Varo,  Simone and Dimastrodonato,  Valeria and Pelucchi,  Emanuele},
  year = {2020},
  month = sep 
}

@article{CavityFeeding_Kaer,
  title = {Microscopic theory of phonon-induced effects on semiconductor quantum dot decay dynamics in cavity QED},
  volume = {86},
  ISSN = {1550-235X},
  DOI = {10.1103/physrevb.86.085302},
  pages = {085302},
  number = {8},
  journal = {Phys. Rev. B},
  publisher = {American Physical Society (APS)},
  author = {Kaer,  P. and Nielsen,  T. R. and Lodahl,  P. and Jauho,  A.-P. and Mørk,  J.},
  year = {2012},
  month = aug 
}

@article{CavityFeeding_Cygorek,
  title = {Nonlinear cavity feeding and unconventional photon statistics in solid-state cavity QED revealed by many-level real-time path-integral calculations},
  author = {Cygorek, M. and Barth, A. M. and Ungar, F. and Vagov, A. and Axt, V. M.},
  journal = {Phys. Rev. B},
  volume = {96},
  issue = {20},
  pages = {201201},
  numpages = {5},
  year = {2017},
  month = {Nov},
  publisher = {American Physical Society},
  doi = {10.1103/PhysRevB.96.201201},
}

@article{orbitalhole,
  title = {Coherent control of a high-orbital hole in a semiconductor quantum dot},
  volume = {18},
  ISSN = {1748-3395},
  DOI = {10.1038/s41565-023-01442-y},
  number = {10},
  journal = {Nat. Nanotechnol.},
  publisher = {Springer Science and Business Media LLC},
author = {Yan, Jun-Yong and Chen, Chen and Zhang, Xiao-Dong and Wang, Yu-Tong and Babin, Hans-Georg and Wieck, Andreas D. and Ludwig, Arne and Meng, Yun and Hu, Xiaolong and Duan, Huali and others},
  year = {2023},
  month = jul,
  pages = {1139–1146}
}

@article{ReappearanceHanschke,
  title = {Experimental Measurement of the Reappearance of Rabi Rotations in Semiconductor Quantum Dots},
author = {Hanschke, Lukas and Bracht, Thomas K. and Sch\"oll, Eva and Bauch, David and Berger, Eva and Kallert, Patricia and Peter, Melina and Garcia, Ailton J. and Covre da Silva, Saimon F. and Manna, Santanu and others},
  journal = {Phys. Rev. Lett.},
  volume = {135},
  issue = {26},
  pages = {263602},
  numpages = {7},
  year = {2025},
  month = {Dec},
  publisher = {American Physical Society},
  doi = {10.1103/s212-43gs}
}

@article{ReappearanceVagov,
  title = {Nonmonotonic Field Dependence of Damping and Reappearance of Rabi Oscillations in Quantum Dots},
  volume = {98},
  ISSN = {1079-7114},
  DOI = {10.1103/physrevlett.98.227403},
  pages = {227403},
  number = {22},
  journal = {    Phys. Rev. Lett.},
  publisher = {American Physical Society (APS)},
  author = {Vagov,  A. and Croitoru,  M. D. and Axt,  V. M. and Kuhn,  T. and Peeters,  F. M.},
  year = {2007},
  month = jun 
}

@article{Wiercinski2023,
  title = {Phonon coupling versus pure dephasing in the photon statistics of cooperative emitters},
  author = {Wiercinski, J. and Gauger, E. M. and Cygorek, M.},
  journal = {Phys. Rev. Res.},
  volume = {5},
  issue = {1},
  pages = {013176},
  numpages = {11},
  year = {2023},
  month = {Mar},
  publisher = {American Physical Society},
  doi = {10.1103/PhysRevResearch.5.013176}
}

@article{zielinski2013fine,
  title={Fine structure of light-hole excitons in nanowire quantum dots},
  author={Zielinski, M},
  journal={Phys. Rev. B},
  volume={88},
  number={11},
  pages={115424},
  year={2013}
}

@article{sallen2009exciton,
  title={Exciton dynamics of a single quantum dot embedded in a nanowire},
  author={Sallen, Gregory and Tribu, Adrien and Aichele, Thomas and Andr{\'e}, R{\'e}gis and Besombes, Lucien and Bougerol, Catherine and Tatarenko, Serge and Kheng, Kuntheak and Poizat, J Ph},
  journal={Phys. Rev. B},
  volume={80},
  number={8},
  pages={085310},
  year={2009},
  publisher={APS}
}

@article{chen2016controlling,
  title={Controlling the exciton energy of a nanowire quantum dot by strain fields},
  author={Chen, Yan and Zadeh, Iman Esmaeil and J{\"o}ns, Klaus D and Fognini, Andreas and Reimer, Michael E and Zhang, Jiaxiang and Dalacu, Dan and Poole, Philip J and Ding, Fei and Zwiller, Val and others},
  journal={Appl. Phys. Lett.},
  pages = {182103},
  doi = {10.1063/1.4948762},
  volume={108},
  number={18},
  year={2016},
  publisher={AIP Publishing}
}

@article{swiderski2017atomistic,
  title={Atomistic theory of excitonic fine structure in InAs/InP nanowire quantum dot molecules},
  author={{\'S}widerski, M and Zieli{\'n}ski, M},
  journal={Phys. Rev. B},
  volume={95},
  number={12},
  pages={125407},
  year={2017},
  publisher={APS}
}

@article{zielinski2025double,
  title={Double nanowire quantum dots and machine learning},
  author={Zieli{\'n}ski, Micha{\l}},
  journal={Scientific Reports},
  volume={15},
  number={1},
  pages={5939},
  year={2025},
  publisher={Nature Publishing Group UK London}
}

@article{swiderski2021electric,
  title={Electric-field control of exciton fine structure in alloyed nanowire quantum dot molecules},
  author={{\'S}widerski, Micha{\l} and Zieli{\'n}ski, Micha{\l}},
  journal={Phys. Rev. B},
  volume={104},
  number={19},
  pages={195411},
  year={2021},
  publisher={APS}
}

@article{swiderski2019electric,
  title={Electric field tuning of excitonic fine-structure splitting in asymmetric InAs/InP nanowire quantum dot molecules},
  author={{\'S}widerski, Micha{\l} and Zieli{\'n}ski, Micha{\l}},
  journal={Phys. Rev. B},
  volume={100},
  number={23},
  pages={235417},
  year={2019},
  publisher={APS}
}
\newpage

\end{document}